\newcommand{\Ls}{L_{\rm s}}
\newcommand{\Ts}{T_{\rm s}}
\newcommand{\vs}{v_{\rm s}}

\newcommand{\vsn}{v_{\rm s,9}}

\newcommand{\tff}{t_{\rm ff}}
\newcommand{\tic}{t_{\rm IC}}
\newcommand{\tdyn}{t_{\rm dyn}}
\newcommand{\tei}{t_{\rm ei}}

\newcommand{\finj}{f_{\rm inj}}
\newcommand{\T}{\tau_T}

\documentclass[twocolumn]{aastex701}
\usepackage{multirow}
\usepackage{graphicx}

\usepackage{amsmath}
\usepackage{makecell}
\usepackage{xcolor}

\usepackage{tabularx,ragged2e}
\renewcommand{\arraystretch}{1.2}
\newcolumntype{Y}{>{\RaggedRight\arraybackslash}X}

\begin{document}

\title{AT2018cow Powered by a Shock in Aspherical Circumstellar Media}

\author[orcid=0000-0002-4327-2634,sname='Govreen-Segal']{Taya Govreen-Segal}
\affiliation{School of Physics and Astronomy, Tel Aviv University, Tel Aviv 6997801, Israel}
\email[show]{taya@govreensegal.com}  

\author[orcid=0000-0002-4534-7089]{Ehud Nakar} 
\affiliation{School of Physics and Astronomy, Tel Aviv University, Tel Aviv 6997801, Israel}
\email{udini@wise.tau.ac.il}

\author{Kenta Hotokezaka}
\affiliation{Research Center for the Early Universe, Graduate School of Science, The University of Tokyo, Bunkyo, Tokyo 113-0033, Japan}
\affiliation{Max Planck Institute for Gravitational Physics (Albert Einstein Institute), Am M\"uhlenberg 1, Potsdam-Golm, 14476, Germany}
\email{kentah@g.ecc.u-tokyo.ac.jp}

\author{Christopher M Irwin}
\affiliation{Research Center for the Early Universe, Graduate School of Science, The University of Tokyo, Bunkyo, Tokyo 113-0033, Japan}
\email{irwincm@g.ecc.u-tokyo.ac.jp}

\author[orcid=0000-0001-9185-5044]{Eliot Quataert}
\affiliation{Department of Astrophysical Sciences, Princeton University,
Princeton, NJ 08544, USA}
\email{quataert@princeton.edu}





\begin{abstract}
We present a quantitative model for the luminous fast blue optical transient AT2018cow in which a shock propagating through an aspherical circumstellar medium (CSM) produces the X-ray and UV/optical/NIR emission. X-rays are emitted from hot post-shock electrons, and soft X-ray photons are reprocessed into optical/UV emission in the cool downstream. This naturally explains two previously puzzling features: (i) the coordinated evolution of the optical and soft X-ray after day 20, (ii) the hard X-ray hump above 10 keV that disappears around day 15 as the Thomson optical depth transitions from $\T \gg1$ to $\T \sim 1$.

Our model is over-constrained, and it quantitatively reproduces the bolometric luminosity evolution, soft X-ray spectrum, and time-dependent soft/hard X-ray and soft X-ray/optical luminosity ratios. It also explains additional puzzles: X-ray fluctuations with  $\sim4-10$ day timescales arise from a global radiative shock instability, while the NIR excess and the apparent receding blackbody radius result from reprocessed X-rays in matter far from thermodynamic equilibrium. The radio is naturally explained as originating from a shock driven by the same ejecta in the more dilute CSM. The light curve steepening after $\sim 40$ days likely indicates the shock reaches the edge of the dense CSM at $\sim {\rm few} \times 10^{15}$ cm. We infer explosion energy $\sim 1-5 \times 10^{50}$ erg, carried by an ejecta at $\sim 0.1c$ and a mass of $0.01-0.05 M_\odot$, in a dense asymmetric CSM with $\sim 0.3 M_\odot$, embedded in a more dilute CSM.
\end{abstract}


\section{Introduction}
Over the last decade, advances in wide-field time-domain surveys have revealed new populations of fast evolving optical transients \citep{Drout2014,Arcavi2016,Tanaka2016,Pursiainen2018,Ho2023,Toshikage2024}. One class of these is the fast blue optical transients (FBOTs), characterized by hot temperatures and rapid decline, and likely originating from extreme supernovae, and the more extreme, luminous fast blue optical transients (LFBOTs), distinguished by peak luminosities $\gtrsim10^{44}\,{\rm erg\,s^{-1}}$. Despite the growing number of detected LFBOTS, these transients exhibit significant diversity in their light curves, spectra, environments, and host-galaxy properties, and it remains unclear whether they form a coherent class or reflect multiple underlying physical mechanisms.

One of the most emblematic members of this emerging population is AT2018cow, discovered by ATLAS \citep{Prentice2018}. AT2018cow was exceptionally bright (peak luminosity $\gtrsim 4\cdot 10^{44} {\rm erg\,s^{-1}}$, nearby (distance $\sim60$ Mpc), and fast-evolving (rise $\sim1$ day, decline timescale of several days). The transient was observed across the electromagnetic spectrum, from radio to hard X-rays \citep[e.g.,][]{Rivera2018,Margutti2019,Ho2019,Perley2019}, yielding one of the richest multi-wavelength data sets for any such event. These observations revealed several remarkable properties: radio and submillimeter data imply a mildly relativistic outflow with velocity $\sim 0.1-0.2\,c$; UV-optical-IR photometry shows an apparently receding blackbody radius (under the assumption that the emission is close to thermodynamic equilibrium); and the X-ray continuum exhibits a hard X-ray hump at early times, followed by large-amplitude variability on timescales of days beginning around day 10.

The breadth of observations associated with AT2018cow has led to a wide range of interpretations. Proposed power sources include: Central engines, such as accreting compact objects or magnetar, often partially covered by external matter which is responsible of the reprocessing ox X-rays into lower-energy emission \citep{Prentice2018,Perley2019,Lyutikov2019,Margutti2019,Piro2020,Uno2023,Li2024,Tsuna2025}, sometimes invoking analogies to tidal disruption events (TDEs) \citep{Kuin2019,Perley2019,Cao2024}; or a supernova in a binary with a black hole \citep{Lazzati2024}. An additional energy source often invoked is a relativistic or mildly relativistic jet \citep{Perley2019,Gottlieb2022}, although  VLBI observations \citep{Bietenholz2020} constrain long-lived outflows to a mean velocity of below 0.5 c at 100 days. Another common component is dense CSM interaction, with various geometries and mass-loss histories, powered by a continuous outflow or an explosion. This is invoked almost in all models to explain the radio emission, and several models also consider this as a source for additional wavelengths \citep{Margutti2019,Fox2019, Pellegrino2022}, such as following a white dwarf merger \citep{Lyutikov2019,Lyutikov2022}, or a pulsational pair instability supernova \citep{Leung2020}.

The goal of this paper is to derive a quantitative model for AT2018cow that explains the multi-wavelength observations with a simple astrophysical setup and a minimal number of free parameters. In our model, an explosion in an aspherical CSM, composed of a dense and dilute component, drives a shock through both components. A fast-cooling shock in the dense CSM produces the X-rays, which are reprocessed in the cooled downstream of the shock to produce UV, optical, and IR emission. The shock in the dilute part of the CSM, produces the radio and sub-mm emission we observe. We show that the model provides a consistent explanation for a wide range of multi-wavelength observations, including the X-ray variability, which originates from a global radiative shock instability, and the IR "excess" which naturally occurs when the X-rays are reprocessed in the downstream of the shock. To the best of our knowledge, this is the first work to model AT2018cow with significantly fewer parameters than observables.  Our model differs from most previous work in that a long-lived central engine is not required to explain AT2018cow's emission on month-timescales and that the time evolution of the source is set largely by the density structure of the CSM and the explosion energy, not by the time-dependent accretion/outflows of a central engine.  This does not preclude a compact object central engine being important for, e.g., the very late time emission.

Although our focus is on AT2018cow, the framework we develop is general and can be applied to other LFBOTs and related fast transients, if they are powered by interaction.

We begin in §\ref{sec:motivation} by summarizing the key observational features of AT2018cow and motivating a shock-interaction interpretation. In \S\ref{sec:hydro}, we constrain the hydrodynamics from several prominent observational features, and motivate the necessity of an aspherical CSM configuration. In §\ref{sec:optical_X-ray}, we develop quantitative models for the X-ray and optical emission produced by such a system, and continue in \S\ref{sec:instability} by discussing the global radiative-shock instability, and its implications for the observed X-ray variability. We apply the model to the multi-wavelength observations of AT2018cow in §\ref{sec:2018cow}, comparing our predictions with observations, and examining additional components. Finally, we discuss the success of our model in explaining different components, and the implications for the energetics and masses of the various components, in §\ref{sec:discussion}, and conclude in §\ref{sec:summary}.

\section{Observational clues and motivation}\label{sec:motivation}
We start by summarizing the main observations:
\begin{itemize}
  \item \textbf{Bolometric Luminosity ($L_{\rm bol}$):} g,r and i band show a peak at $\sim\!1-2$ days after explosion. Multi-wavelength observations start 3 days after explosion, and imply $L_{bol}(t=3\rm ~d)\simeq 4\cdot 10^{44} \rm ~erg/s$, after which the bolometric light curve declines roughly as $L_{\rm bol}\propto t^{-2}$ until $\sim\!40$ days, followed by a steepening \citep{Perley2019,Margutti2019,Ho2019}.
  \item \textbf{Optical light curve:} peaks at $\sim\!1$--$2$ days. The optical then declines but remains a substantial fraction of $L_{\rm bol}$ ($\gtrsim 50\%$) throughout. At $t\simeq3$ days, the blackbody radius is $\simeq8\cdot10^{14}\,{\rm cm}$, suggesting a mean photospheric velocity  $\sim0.1$-$0.2\,c$ at that epoch \citep{Perley2019,Margutti2019,Ho2019}. 
  \item \textbf{Soft X-ray (0.5--10 keV):} the luminosity spectrum is $L_\nu\propto \nu^{-1/2}$ throughout; and at day $\sim 8$ shows a feature at $\sim$ 8 keV with a width of $\sim 2-4$ keV; the light curve is approximately flat until day~20, after which it tracks the optical luminosity with $L_{\rm {x,soft}}\approx L_{\rm opt}$. \citep{Rivera2018,Perley2019,Margutti2019,Ho2019}, see also Fig. 9 in \cite{LeBaron2025}. 
  \item \textbf{Hard X-ray ($>$10 keV):} a distinct high-energy hump persists until $\sim$day~15. At later times, the hard X-ray spectrum follows the soft X-ray continuum at least till $\sim$day~35 \citep{Margutti2019}.
  \item \textbf{X-ray fluctuations}: From $\sim$day 10, the X-ray luminosity (hard and soft) starts fluctuating by up to a factor of $\sim4$ in flux. The fluctuation timescale increases slowly with time (sub-linearly) from about 4 days on day 10 to about 10 days around day 50 \citep{Margutti2019,Kuin2019}.
  \item \textbf{Radio}: Observations suggest a synchrotron origin. Applying synchrotron self-absorption theory implies a mean velocity of the radio-emitting region $\sim 0.1-0.2c$ at 30-130 days \citep{Margutti2019,Ho2019}. 
  \item \textbf{Sub-millimeter:} Seems to originate from the same source of the radio emission, and shows a consistent velocity also at day 23, although its spectrum is steep compared to a regular optically thin synchrotron spectrum \citep{Ho2019}. 
  \item \textbf{Spectral lines}: Until day 8, the spectrum is featureless except for a possible broad absorption feature that corresponds to about 0.1c. After that, broad emission features appear with widths of $\sim 10,000\,{\rm km/s}$, and become narrower with time, reaching $\sim 4000\,{\rm km/s}$ after day 20. At around day 20, narrow hydrogen and helium lines start appearing \citep{Perley2019,Margutti2019}. 
  \item \textbf{Optical polarization}: peaks at $\simeq 7\%$ at $\sim 6$ days, implying asymmetry \citep{Maund2023}.
  \item \textbf{Late time emission}: HST Observations 700-2000 days after the event show slowly declining emission, and place a lower limit on the luminosity of $\sim 10^{39}\,{\rm erg/s}$ for these epochs, and blackbody temperatures in the UV, with high uncertainty since the peak is not observed \citep{Sun2022,Chen2023,Inkenhaag2023,Inkenhaag2025}. \cite{Migliori2024} observe late time X-rays, and report $\sim 2\cdot 10^{39}\,{\rm ergs/s}$ at day 218, and $\sim 4\cdot 10^{38}\,{\rm erg/s}$ at day 1350, with a soft spectrum in the $0.2-4.5$\,keV range, though it is unclear whether the later epoch is due to the transient or the host galaxy. 
    \item \textbf{NIR emission} is in excess of the blackbody fit to the optical, and 3-5 times fainter than the X-ray emission throughout. The excess over the blackbody fit increases significantly after day 20  (\cite{Perley2019}, see also \cite{LeBaron2025} fig. 9). 
  \item \textbf{X-ray quasi-periodic oscillations (QPOs)}: \cite{Pasham2022} suggest a $\sim 200$ Hz QPO with a 3.7$\sigma$ significance level, and \cite{Zhang2022} suggest a $\sim 4$ mHz QPO with 2.2$\sigma$ significance. We treat these as tentative, since their significance is not high for posteriori statistics. 
\end{itemize}
Below, we start by focusing on the X-ray and optical emission, and discuss two striking features that stand out in these observations, and which motivate our model.
\paragraph{\#1: Coordinated optical--X-ray evolution.}
Starting around day~20, the soft X-rays (0.5-10 keV) and the optical\footnote{Throughout, “optical” includes NIR and UV.} evolve together, maintaining $L_{\rm X,soft}\simeq L_{\rm opt}$, strongly suggesting that they originate from the same source. A simple way to get similar luminosities is if the optical is reprocessed X-ray emission.

Two reprocessing geometries can yield $L_{\rm opt}\sim L_{\rm X}$: (i) an X-ray source obscured by a reprocessing medium with covering factor $\sim\!1/2$, such as suggested by \cite{Margutti2019}, and (ii) an X-ray source that lies between the observer and the reprocessing region,  so that roughly half the source photons reach the observer and half are reprocessed. A fast-cooling collisionless shock can supply such a geometry, as the hot electrons radiate in X-ray, while the cool dense downstream efficiently absorbs X-rays and re-emits in the optical/UV. 

\paragraph{\#2: Hard X-ray hump and its disappearance.}
The hard X-ray excess above 10\,keV until $\sim$day~15 resembles a reflection/Compton hump. Its disappearance at around day 15, is difficult to explain with an obscured X-ray source. In a shock bounded by a reprocessing downstream and a highly ionized upstream with Thomson optical depth $\T$, soft X-rays are attenuated by a factor $\sim 1/\T$ while $>\!10$ keV photons are scattered in the downstream and unattenuated, naturally producing the "hump" when $\T\gg1$, which decreases and joins the soft X-ray continuum as $\T$ approaches unity \citep{GS2025}.

\paragraph{Main idea}
These properties suggest that the X-ray and optical emission are powered by a shock in dense CSM. In this scenario, the early optical peak at $t\sim1$-$3$ days is due to a shock breakout, which is followed by a collisionless shock propagating through a CSM that is initially optically thick to Thomson scattering. The shock is fast-cooling at all times, either via inverse-Compton or free-free, and the cooling is fast enough that the downstream efficiently absorbs and reprocesses soft X-rays into optical emission.  In this scenario, while $\T>1$, soft X-ray photons have a probability of $\sim \frac{1}{\T}$ to reach the observer without being reprocessed in the downstream, leading to an attenuation of $\sim \frac{1}{\T}$ of soft X-rays relative to hard X-rays, and to the inequality $L_{opt}\ge L_{x,soft}/\T$, which turns to equality if free-free dominates the cooling. Once $\T <1$, about half of the X-ray emission reaches the observer directly, while the other half reaches the cool downstream. Therefore, the X-ray should not have a prominent hard X-ray hump, and $L_{opt}\gtrsim L_{x,soft}$, where again, equality is attained if free-free dominates the emission. 

Below, we explore whether this model, of a single fast-cooling shock in CSM can explain the entire set of optical and X-ray observations of AT2018cow. The observables that we attempt to explain are: (i) $t_{\rm bo}\simeq1$--$2$~d; (ii) $v_{\rm bo}\simeq0.1$--$0.2\,c$, as inferred from the optical photosphere at day 3 and the broad spectral features at early times; (iii) $L_{\rm bol}(3\,{\rm d})\simeq4\cdot10^{44}\,{\rm erg\,s^{-1}}$; (iv) $L_{\rm bol}\propto t^{-2}$ from 3 days to $\sim40$~d; (v) the observed $L_{x,soft}/L_{opt}$ at all times; (vi) the observed  $L_{x,soft}/L_{x,hard}$ at all times; (vii) X-ray spectrum; and 
(viii) X-ray fluctuation (qualitatively). As we discuss next, our model contains only four parameters that fully constrain the shock hydrodynamics. We allow these parameters to vary in the limited range allowed by four observables, and check if there is a solution that explains the entire set of observations.

\begin{figure}
    \centering
    \includegraphics[width = \columnwidth]{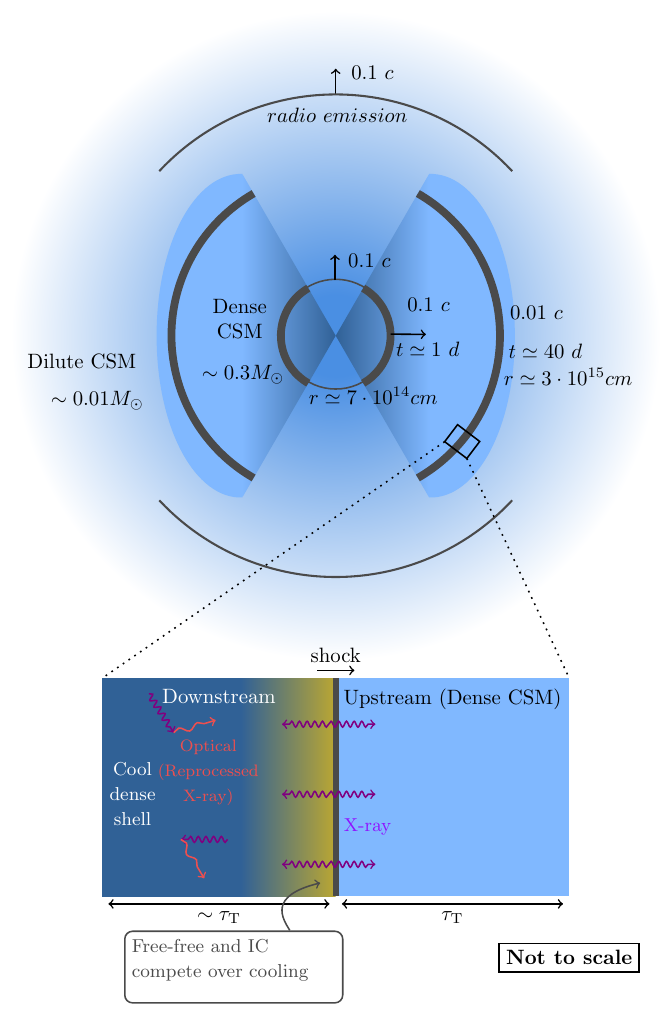}
    \caption{A schematic sketch of our model, showing on the top, the asymmetrical dense CSM embedded in a more dilute CSM. A single explosion launches a shock in both components; the X-ray and optical originate from a shock in the dense CSM, which decelerates as it collects mass, while the radio is emitted by the constant velocity shock in the dilute CSM.}
    \label{fig:sketch}
\end{figure}

Figure \ref{fig:sketch} demonstrates the setup we have in mind throughout this paper. It serves as an example for a CSM possible configuration, though many different asymmetrical configurations can serve our model equally well.

\section{Hydrodynamics}\label{sec:hydro}
Consider a shock propagating in CSM with a power-law CSM density profile:
\begin{equation}\label{eq:rho}
  \rho(r)=\rho_{\rm bo}\,\left(\frac{r}{r_{\rm bo}}\right)^{-s},\qquad
\end{equation}
such that the shock velocity is decreasing as a power-law in time:
\begin{equation}
      v_s(t)=v_{\rm bo}\,\left(\frac{t}{t_{\rm bo}}\right)^{-k}.
\end{equation}
With these definitions, the system is fully specified by four parameters: the breakout time $t_{\rm bo}$, the breakout velocity $v_{\rm bo} \equiv v_s(t_{\rm bo})$, and the two power-law indices $s$ and $k$.  The breakout time and velocity are directly constrained observationally from the breakout time and photospheric velocity, and $\rho_{\rm bo} \equiv \rho(r_{\rm bo})$ is tied to the breakout optical depth $\tau_{bo}=\frac{c}{v_{bo}}$ via $\tau_{\rm bo}\simeq \frac{\sigma_{T}}{m_{p}}\frac{\rho_{bo}r_{bo}}{1-s},$
where $r_{\rm bo}= v_{bo}t_{bo}/(1-k)$ is the breakout radius, implying the density at breakout is related to $t_{bo}$ and $v_{bo}$ by: 
\begin{equation}\label{eq:rho_bo}
    \rho_{bo}=\frac{m_p c (s-1)(1-k)}{\sigma_T \cdot}t_{bo}^{-1}v_{bo}^{-2}.
\end{equation}

$s$ and $k$ can be found from the following two observational constraints. The first is the Thomson optical depth evolution, where in our model the optical depth at the $t_{bo}=1-2\rm~d$ is $\T=\frac{c}{v_s}\simeq 5-10$ and the optical depth at the time that the hard X-ray hump disappears is about unity,  $t(\T{=}1)\simeq15-20\,\rm d$. The evolution of the optical depth as a function of the model parameters is 
\begin{equation}\label{eq:tau}
  \T = \frac{c}{v_{bo}} \left(\frac{t}{t_{bo}}\right)^{-b},
\end{equation} 
where we defined \begin{equation}\label{eq:b_def}
    b\equiv (1-k)(s-1),
\end{equation}
to simplify the expressions throughout this paper. 
Requiring that the optical depth drops from 5-10 to unity between days 1-2 to days 15-20 implies $b=0.5-1.15$.
The second constraint is the observed luminosity evolution, under the assumption of a fast cooling shock:
\begin{equation}\label{eq:Ls}
  L_{\rm bol}(t)\;\simeq\; L_s \simeq f_\Omega\cdot 2\pi\,\rho_{us}\,r^2\,v^3=L_{bo} \left(\frac{t}{t_{bo}}\right)^{1-b-4k}.
\end{equation}
where $L_s$ is the total thermal energy flux through the shock (i.e., half the total energy flux, since about half the shock energy remains kinetic). $f_{\Omega}$ is a geometrical factor that denotes the solid-angle covering factor of the dense CSM $f_\Omega\equiv \Omega/4\pi$, which in our model cannot be much smaller than unity, otherwise the outflow engulfs the CSM.

Combining the observed decline of the bolometric luminosity, $L_{\rm bol}\propto t^{-2}$ and of $\T$, we obtain
\begin{equation}\label{eq:k_s}
  s \simeq 2.4-3.1,\qquad
  k=\frac{4-s}{5-s}\simeq 0.47-0.62.
\end{equation} 
 Importantly, this implies a shock decelerating faster than any shock in spherical geometry with the relevant density profile. The most steeply decelerating shock in spherical symmetry is the momentum-conserving \emph{snowplow} solution, expected for a radiative shock in spherical media, for which: $ k_{\rm sp}=\frac{3-s}{\,4-s\,}$. The reason for this is the steep decline of the bolometric luminosity, which is faster than any solution of a fast-cooling spherically symmetric shock.

The larger $k$ inferred here suggests that energy and momentum are lost from the shock as it propagates. One way to accomplish this is in a non-spherical configuration in which energy and momentum leak out laterally and are lost from the system. A second, related effect that can explain the fast luminosity decline is a non-constant opening angle of the medium, so $f_\Omega$ decreases with the radius\footnote{The shock velocity in either case will not necessarily decrease as a power-law in time. However, it can be approximated by a power-law over a single order of magnitude in time, as is considered here.}. Given that we do not expect $f_\Omega$ to be much smaller than unity, we assume, for simplicity, that it is constant. Thus, as a possible configuration, we propose a dense CSM that is focused around the equator, and has a fixed half-opening angle $\theta_{\rm CSM}$, so  $f_\Omega = \sin\theta_{\rm CSM}$.
 
As our model is accurate to within a factor of a few, we cannot measure $f_{\Omega}$, but only make sure our model is consistent with it, by verifying that $L_{bol}(t=\rm 3d)\approx4\cdot 10^{44} {\rm~ erg/s} = L_{bo}\left(\frac{3~\rm d}{t_{bo}}\right)^{-2}$, with $f_{\Omega}\simeq 0.5-1$.
Indeed, plugging in parameters within the acceptable range, we find that $f_\Omega\simeq 0.5-1$ favors $t_{bo}\approx 1\rm ~d$ and $v_{bo}\approx 0.1c$ but is broadly consistent with most of the phase space allowed by the observations. This serves as a first consistency test for our model. From here on, we fix $f_\Omega=1$ for our analysis.

A dense CSM that covers only a part of the solid angle, in addition to explaining the enhanced deceleration, also accommodates: (i) a radio-emitting shock that propagates through lower-density regions (where there is no thick CSM), and does not decelerate significantly between breakout and $\sim 130$ days, so that it is consistent with the values inferred from synchrotron self-absorption (SSA) \citep{Margutti2019,Ho2019}, and (ii) early optical polarization indicating asphericity \citep{Maund2023}. 

With the hydrodynamic parameters constrained, we can now test whether a shock with these parameters is also consistent with the range of additional observables (v-viii in the list given at the end of the previous section). To achieve this, we begin by deriving a phase space that describes the system (e.g., dominant cooling process, optical depth, etc.), and then find the emission in each phase, constructing light curves that can be compared to observations.

\section{Emission and reprocessing}
In the interaction-powered picture, the observed radiation is a composition of (i) free-free X-ray emission from the hot immediate downstream (ii) IR/optical/UV radiation produced by reprocessing of the X-rays in the cold downstream, and (iii) additional energy added to the soft radiation field as it cools the hot post-shock electrons via IC\footnote{Note that the IC cooling only contributes to the IR/optical/UV, since photons have a high probability of being reprocessed between every scatter in the hot region, and thus cannot be shifted into the X-ray band by repeated IC scattering.}. To understand the emission from these three processes, we first need to determine the process that dominates the cooling. The relevant timescales can be expressed in terms of $\vs,\T,\rho$, and the dependence on $\rho$ drops out when equating any two timescales, making it convenient to plot the state of the system on a phase space of $\vs$-$\T$. Below, we discuss the relevant cooling regimes, provide scalings for the luminosity in each band, and assemble the expected light curves.

\subsection{Characteristic timescales}
The important timescales are the dynamical, free-free, inverse-Compton and Coulomb coupling timescales. While adiabatic cooling cools both species, inverse Compton and free-free cooling cool the electrons, and require $T_i/T_e$ electron cooling times in order for the ions to be cooled, and their energy to be radiated by the electrons. When they are in equipartition, as is relevant for a large part of the phase space, a single electron cooling time is sufficient for cooling the ions. 

We thus write the timescales for cooling the ions:
\begin{align}\label{eq:tdyn}
\tdyn &= \frac{r_s}{\vs}
      = 13\,{\rm d}\,(s-1)\,\T~ \rho_{us,-15}^{-1}\,\vsn^{-1},\\
\tff  &= 77\,{\rm d~}\,T_{i,9}T_{e,9}^{-1/2}\,\rho_{-15}^{-1}, \label{eq:tff}
\end{align}
where $\rho_{us}$ is the density in the immediate upstream of the shock, $\rho$ is the local density (immediately behind the shock $\rho=4\rho_{us}$, farther downstream $\rho$ changes according to the pressure and the temperature), $T_e$ is the local electron temperature and  $T_i$ is the local ion temperature. $Q_x\equiv Q/10^x$ in cgs units, i.e., $\vsn\equiv v_{\rm s}/10^{9}\,{\rm cm\,s^{-1}}$, $\rho_{us,-15}\equiv \rho_{\rm us}/10^{-15}\,{\rm gr\cdot cm^{-3}}$, etc. 
To find the IC cooling time, we note that the origin of the soft photon field cooling the hot electrons by IC is itself generated by the shock. Since the IC cooling time depends on the energy density in the soft photon field, it is useful to parametrize the soft photon energy density in terms of the shock luminosity. We thus define the ratio between the actual radiation energy density of the soft photons and the energy density of radiation in the case of a fast cooling optically thin shock (i.e., the entire shock energy flux is converted to photons that stream freely to the observer):
\begin{equation}
\finj \equiv \frac{u_\gamma(r_s)\,4\pi r_s^2 c}{\Ls}.
\end{equation}
In terms of $\finj$, the value of which we find later, the IC cooling time is given by:
\begin{equation}\label{eq:tic}
    \tic  =  12\,{\rm d~} \rho_{us,-15}^{-1}\,\vsn^{-3}\,\finj^{-1}T_{e,9}^{-1}T_{i,9}.
\end{equation}

The Coulomb coupling between ions and electrons, which proceeds on
\begin{equation}\label{eq:tei}
\tei = 6\,{\rm d~}\,T_{e,9}^{3/2}\rho_{-15}^{-1}\,\bar{Z}^{-2}\,\Lambda_{e,25}^{-1}.
\end{equation}
Here, $\bar{Z}^2 \equiv \frac{\sum_i n_i Z_i^2}{\sum_i n_i}$ is the mean ion charge (with $Z_i$ the ionic charge state, $n_i$ the density of the corresponding ion); for a fully ionized, solar metallicity gas $\bar{Z}\simeq 1.1$, and $\Lambda_e$, is the Coulomb logarithm, which has a typical value of 25. Note that $\tei(T_e=T_i)$ (i.e., replacing $T_e$ with $T_i$) is the timescale for the electrons to reach equipartition with the ions, while $\tei\cdot \frac{T_e}{T_i-T_e}$ is the timescale for electrons to double their temperature due to heating from the ions, which for $T_e\ll T_i$ can be approximated as $\tei\cdot \frac{T_e}{T_i}$. 

The relative order of these timescales sets the state of the system. We defer a complete treatment of the different regimes to future work, and focus here only on the fast-cooling regimes, which are of interest for AT2018cow: fast free-free and fast IC cooling. We ignore here any contribution of non-thermal electrons, which are probably accelerated by the shock, as this contribution is expected to be negligible in the Optical and X-ray bands.

\paragraph{(A) Fast free-free cooling {\rm (}$\tff<\tdyn,\tic${\rm )}}
Under the assumption of electron-ion equipartition (to be verified shortly), the immediate downstream temperature, which is set by the shock jump conditions, is:
\begin{equation}\label{eq:Ts}
k_B\Ts \simeq 100~{\rm keV}\,\vsn^2.
\end{equation}
Thus, from Eq. \ref{eq:tdyn} \& \ref{eq:tff} we find that $\tff<\tdyn$ for:
\begin{equation}\label{eq:ff_fast_condition}
\T \gtrsim  1.6 (s-1)^{-1}\,\vsn^{2}. 
\end{equation}
 We verify that in this regime, $\tei(T_e=T_i)<\tff,\tic$, implying that our assumption of electron-ion equipartition is justified and that for our purposes $T_e\sim T_i$ at all times, regardless of the (unknown) coupling of electrons and protons within the shock transition layer.

The free-free emission spectrum in this regime is made up of many electron populations of different temperatures, cooling by radiating their energy. Asymptotically (for $h\nu\ll kT_{s}$), the contributions from the various electron populations add a logarithmic correction. However, this limit is never approached, since as we will see, free-free dominates the emission only for $1{\rm ~keV}\lesssim T_e\lesssim60{\rm ~keV}$, where above the upper limit IC dominates and below the lower limit line cooling dominates. An integral solution for this spectrum was derived by \cite{Wasserman2025}. In appendix \ref{App: ff_spec} we re-derive this spectrum, and show that for $h\nu \gtrsim 0.02~ k_B\Ts$, it is well approximated by: 
\begin{equation}\label{eq:ff_spectrum}
L_\nu \propto 
\nu^{-1/2}\,\exp\left(-h\nu/k_B \Ts\right).
\end{equation}
In principle, to check whether the shock is fast cooling enough for the downstream to efficiently reprocess the soft X-rays, we need the cooling length to be smaller than a single Thomson scattering mean free-path, e.g., $\tff<\tdyn \cdot \min(\T^{-1},1)$. However, we find that free-free is the dominant process only for $\T \lesssim1$, for which this criterion coincides with fast cooling. 
\paragraph{(B) Fast IC cooling with reprocessing.}
When IC is the dominant cooling process, nearly the entire shock luminosity is emitted in soft photons\footnote{In principle, for IC to be efficient, the seed radiation field must be large enough to accommodate cooling the shock with inverse Compton. As we will see, all observations of AT2018cow fall in the regime where free-free cooling is efficient, or nearly so, and this is not a concern.}. To derive $\finj$, we consider that for $\T>1$, every photon is scattered 
$\sim \T^2$ times on its way out of the media, which means it scatters $\sim \T^2-(\T-1)^2=2\T-1\sim \T$ times through the shock radius, increasing the local energy density by a factor of $\sim \T$, while for $\T\ll 1$, half the emitted photons are scattered backwards, and will pass through the shock again, and half escape to the observer, so we adopt
\begin{equation}\label{eq:finj_tau}
\finj \simeq \max(\T,0.5).
\end{equation}
We must also consider whether the electrons and ions are in equipartition. In the part of phase space that $\tei(T_e\simeq T_i)<\tic(T_e\simeq T_i)$, we assume $T_e\simeq T_i$ and find where there is fast IC cooling, by comparing $\tic$ and $\tdyn$. Where $\tic<\tei<\tdyn$, the electrons and ions are not in equipartition\footnote{Even if electrons are coupled to ions in the shock, if the electrons cool fast and $\tic<\tei$, they will reach a temperature $T_e\ll T_i$ where coulomb heating is equal to IC cooling.}, $T_e\ll T_i$. $T_e$ is set by the balance of electron cooling by IC,  which has a typical timescale $\tic\cdot \frac{T_e}{T_i}$, and the time it takes the electrons to double their temperature by electron-ion interaction, $\simeq\tei \frac{T_e}{T_i}$. Equating the two, we find:
\begin{equation}\label{eq:Ticei}
    k_B T_{e,IC=ei}=270{\rm ~keV}\,\T^{-2/5}\vsn^{-2/5}\Lambda_{e,25}^{2/5}\bar{Z}^{4/5}.
\end{equation}
Using this expression for the electron temperature, and $T_i\simeq \Ts$, we find that there is fast IC cooling ($\tic\le\tdyn$) for:
\begin{equation}\label{eq:IC_fast_condition}
\T \gtrsim 
\begin{cases}
1\,\vsn^{-1} & v_s<v_s^{\rm(eq)}\\[6pt]
\vsn^{1/4}\Lambda_{e,25}^{-1/4}\,\bar{Z}^{-1/2}\, & v_s>v_s^{\rm(eq)}
\end{cases}~'
\end{equation}
where the equipartition threshold is for:
\begin{equation}
v_s^{\rm(eq)} \simeq 1.3\times 10^9 {\rm~cm~s^{-1}~}\Lambda_{e,25}^{1/6}\,\bar{Z}^{1/3}\T^{-1/6},
\end{equation}
i.e., for $\vs\le \vs^{\rm(eq)}$ the electrons and ions are fully coupled so  $T_e\simeq T_i$ and for $v_S>\vsn^{\rm(eq)}$ they are not and $T_e\ll T_i$. 
Using this transition, we find that fast IC cooling takes place only for $\T\gtrsim0.5$, which is why we approximated $\finj\simeq\T$ in all equations above.   
Lastly, we must verify that the cooling behind the shock front is fast enough that reprocessing occurs within a single Thomson scattering length from the shock front. We investigate this in Appendix \ref{App: reprocessing}, and find that for $T_e\simeq T_i$ this is always the case, and that for $T_e\ll T_i$, there is efficient reprocessing in the downstream for:
\begin{equation}
    \T\gtrsim 0.3 \vsn^{2}. 
\end{equation}

\begin{figure}
\centering
\includegraphics[width=\columnwidth]{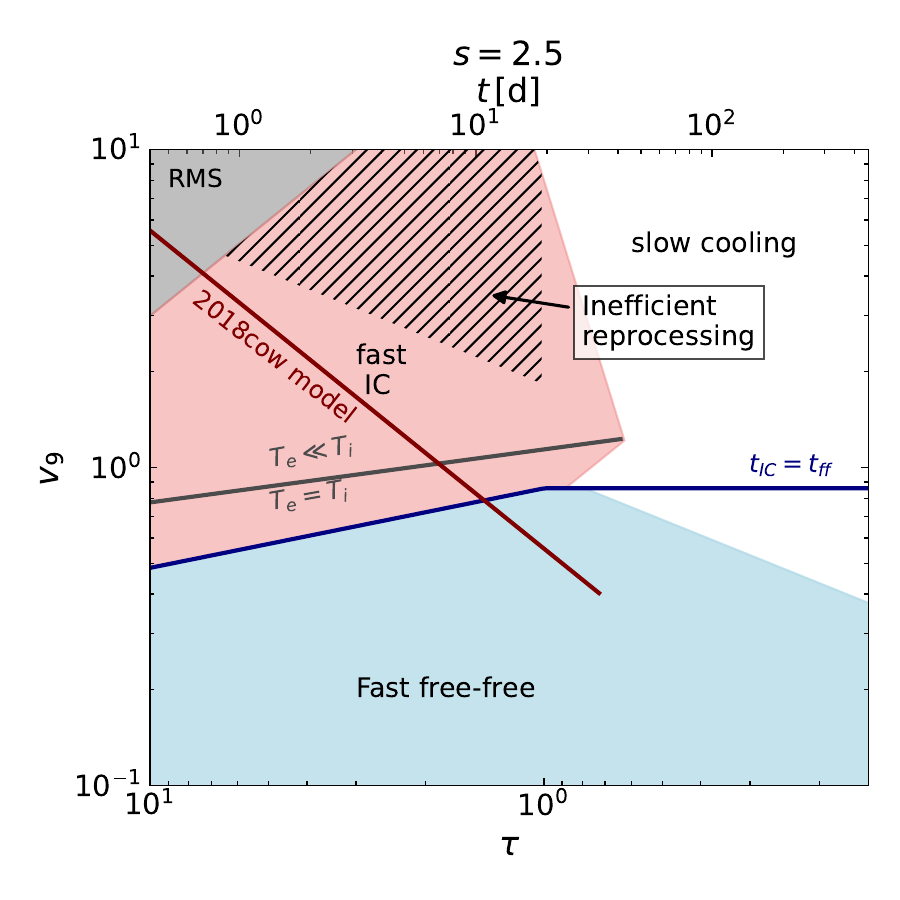}
\caption{Phase space of the various regimes in the $(\T,\vs)$ plane, for $s=2.5$. The red line shows our canonical model for AT2018cow ($t_{bo}=0.7\rm~d$, $v_{bo}=0.13~c$, $s=2.5$, $k=0.6$), and the top axis shows the time corresponding to this model. While the phase space itself only weakly depends on $s$ and is independent of all other parameters, the time depends on all model parameters, and is specific for our canonical model.}
\label{fig:phase_space}
\end{figure}

The IC and free-free fast-cooling regions overlap substantially. The transition between the two is at a velocity of:
\begin{equation}
    v_{IC=ff} \simeq 8.6\cdot 10^8{\rm~ cm~s^{-1}}\cdot \min(\T^{-1/4},1)~,\label{eq:v_tr}
\end{equation}
where for higher shock velocities IC dominates, and for lower velocities free-free dominates. The different regimes are plotted in Figure \ref{fig:phase_space}.

\subsection{Optical and X-ray emission}\label{sec:optical_X-ray}
With the different regimes mapped out, we can now derive the emission in the relevant regimes. The emission is shaped by the shock power, by the dominant cooling process, and by reprocessing. The competition between free-free and IC over cooling the shock sets the intrinsic fraction of the shock energy initially radiated in X-rays. Part of this emission is then reprocessed into optical in the downstream. Assuming the cold region of the downstream is less than a single scattering length away from the shock front (an assumption we will later verify is indeed valid for the parameters of AT2018cow), the vast majority of soft X-ray photons that move away from the shock are reprocessed to optical/UV. The probability of a soft X-ray photon being emitted from the shocked region into the upstream to reach the observer without returning to the shocked region even once, and being reprocessed, is $\sim \min\left(1,\frac{1}{\T}\right)$. Let $L_{x}^{\rm(intr)}$ be the total \emph{intrinsic} X-ray luminosity emerging from the immediate downstream \emph{before} reprocessing, which we will soon derive based on the cooling process. Soft X-rays ($h\nu<$ 10 keV) are reprocessed if they reach the cool layer in the downstream\footnote{Assuming the column density of the reprocessing region has a column density $\gtrsim 10^{24}Z_\odot/Z {\rm~cm^{2}}$, where $Z$ is the gas metallicity,  so it can reprocess photons up to 10 keV, as we will see is valid for AT2018cow. As the column density drops below this value, the reprocessing region becomes transparent to lower energy X-rays and cannot reprocess them, for example, at a column density $\sim6\cdot 10^{22}Z_\odot/Z{\rm ~cm^2}$, only photons below 3 keV will be reprocessed.}, while hard X-rays are reflected by the reprocessing region and escape essentially unattenuated. We therefore write:
\begin{align}
L_{x,\rm soft} &\simeq \frac{L_{x}^{\rm(intr)}(0.3 \le h\nu \le 10 \rm~ keV)}{\max(1,\T)} , \label{eq:soft_rule}\\
L_{x,\rm hard} &\simeq L_{x}^{\rm(intr)}(h\nu \ge 10 \rm ~keV)\label{eq:hard_rule},
\end{align}
where the $1/\max(1,\T)$ factor encodes the soft-photon escape probability.

Using Eq.~\eqref{eq:Ts} and the fast-cooled free-free spectrum of Eq.~\eqref{eq:ff_spectrum}, we can derive the fraction of X-ray emission in the soft X-ray band, which we denote $f_{\rm soft}\equiv\dfrac{L_{x}^{intr}(0.3\le h\nu\le10 \rm~keV)}{L_{x}^{intr}}$:
\begin{equation}\label{eq:f_soft}
\begin{aligned}
    f_{\rm soft}(t) &\simeq 
    \begin{cases}
    \left(\frac{10~\rm keV}{\Ts}\right)^{1/2}, & \Ts > 10~\rm keV, \\
    1 - \left(\frac{1~\rm keV}{\Ts}\right)^{1/2}, & 1~\rm keV \le \Ts \le 10~\rm keV, \\
    0, & \Ts \le 1~\rm keV,
    \end{cases} \\
    &\quad \text{or equivalently,} \\
    &\simeq 
    \begin{cases}
    0.30 \vsn^{-1}, & \vsn > 0.3, \\
    1 - 0.1 \vsn^{-1}, & 0.1 < \vsn < 0.3, \\
    0, & \vsn \le 0.1.
    \end{cases}
\end{aligned}
\end{equation}

and, correspondingly, the hard X-ray fraction gives
\begin{equation}
    L_{x,\rm hard} \simeq L_{x}^{\rm intr}\cdot \max\!\left(1-\dfrac{0.30}{\vsn},\,0\right). \label{eq:Lx_hard_ff}
\end{equation}
With these, we can now derive $L_{x,intr}$ in the fast free-free and fast inverse Compton regimes, so that the light-curve can be assembled. 

\paragraph{Fast free-free regime}
Most of the shock luminosity is emitted in X-rays, $L_{x}^{\rm intr}\simeq L_s$, to within a factor of 2.  

\paragraph{Fast IC cooling}
In the fast IC regime, electrons cool first by IC and then, farther downstream, transition to free–free. The IC$\to$ff transition temperature follows from $\tff=\tic$, evaluated at constant pressure set by the shock front\footnote{In the downstream, if the cooling is fast enough, the pressure doesn't change considerably, see Appendix \ref{App:thin_shell}. This assumption does not consider the potential effect of the pressure of the non-thermal ions. If non-thermal ion pressure can exceed equipartition, the relevant region would be at constant density rather than constant pressure. We find that as the non-thermal ions carry less than 10\%, this change does not influence the expected emission in the observed regimes, though it should be further studied at higher shock velocities.}:
\begin{equation}\label{eq:Ticff}
k_B T_{IC=ff} \simeq 66~{\rm keV}\,\left(\vsn\max(\T,1)\right)^{-2/3}.
\end{equation}
The intrinsic X-ray luminosity from the electron populations that have cooled enough that they are emitting primarily in free-free is:
\begin{align}\label{eq: Lx_intr_IC}
L_{x,2}^{\rm intr}&\simeq \frac{T_{ff=IC}}{\Ts}\,\Ls.
\end{align}
For higher temperature electrons for which IC is the dominant cooling process, the free-free emission can be estimated, assuming the electron temperature is given by Eq. \eqref{eq:Ticei}:
\begin{equation}
    L_{x,1}^{\rm intr}\simeq \Ls\left(1-\frac{T_{IC=ff}}{\Ts}\right)\frac{\tic\cdot \Ts/T_{IC=ei}}{\tff}.
\end{equation}
Comparing the two, we find that $L_{x,2}\gtrsim L_{x,1}$ in the fast IC regime. For simplicity, neglect the contribution of $L_{x,1}$, and approximate $L_{x}^{\rm intr}\simeq L_{x,2}^{\rm intr}$.

For both regimes, we can insert $L_{x}^{\rm(intr)}$ into Eqs.~\eqref{eq:soft_rule}–\eqref{eq:hard_rule} together with $f_{\rm soft}$ from Eq.~\eqref{eq:f_soft} to obtain $L_{x,\rm soft}$, $L_{x,\rm hard}$, and $L_{\rm opt}$. We do so in the following section and express our results in terms of measurable light-curve parameters. Here, we summarize by presenting the predicted soft X-ray/optical and hard/soft X-ray ratios in Table \ref{tab:ratios_expr}. These provide a useful understanding of the system that is roughly independent of the system's hydrodynamics, and can help read $\T$ and $v_s$ directly from observations. This table assumes a significant fraction of the intrinsic X-ray emission is in the hard X-ray band. This assumption is valid for $\vs\gtrsim 4\cdot 10^8 {\rm~cm/s}$, such that the shock temperature is $\gg10 {\rm ~keV}$, at lower temperatures the corrections laid out in equation \eqref{eq:f_soft} must be included.
\begin{table}
    \centering
\begin{tabular}{c|c c}
\hline 
Regime & $L_{x,soft}/L_{opt}$ & $L_{x,hard}/L_{x,soft}$\tabularnewline
\hline 
\hline 
fast IC ($\T\gtrsim1)$ & $\frac{1}{\T}\cdot\frac{(T_{ff=IC} \cdot 10 {\rm ~keV})^{1/2}}{\Ts}$ & $\T\cdot\left(\frac{T_{ff=IC}}{10\,\rm keV}\right)^{\frac{1}{2}}$\tabularnewline
\hline 
fast free-free, $\T>1$ & $\T$ & $\T\cdot\left(\frac{T_{s}}{10\,\rm keV}\right)^{\frac{1}{2}}$\tabularnewline
\hline 
fast free-free, $\T\lesssim1$ & $1$ & $\left(\frac{T_{s}}{10\,\rm keV}\right)^{\frac{1}{2}}$\tabularnewline
\hline 
\end{tabular}
    \caption{We present the ratios between the soft X-ray and optical, and between the hard and soft X-ray. These are shaped by the reprocessing as well as the shock cooling processes, and directly relate to $\T$ and $v_s$, through the dependence of the shock temperature (eq. \ref{eq:Ts}) and $T_{ff=IC}$ (eq. \ref{eq:Ticff}) on these quantities.
    Note that these expressions are approximate. Specifically, they assume that $T_s,T_{ff=IC}>10 {\rm ~keV}$ (i.e., $v_s\gtrsim 4\cdot 10^8 {\rm ~cm/s}$), and assume the shock emission is observable up to the cutoff temperature. $L_{x,soft}/L_{opt}$ does not account for optical contributions from IC (in the free-free dominated regime) and line cooling emission. Here and in all other tables, we use $k_B=1$.}
    \label{tab:ratios_expr}
\end{table}

\subsection{Light-curves}\label{sec:lcs_final}
In this section, we write the lightcurve in a manner that is more accessible, in terms of:
$\{L_{\rm bo},\,t_{\rm bo},\,s,\,k\}$.
We express $v$ and $\T$ in terms of these four parameters, and use them to rewrite the expressions for the light curve presented in the previous section.
\begin{equation}
    v_{bo} = 0.1c\cdot\left(\frac{L_{bo,45}}{f_{\Omega}t_{bo,d}}\frac{\left(1-k\right)}{\left(s-1\right)}\right)^{\frac{1}{3}}.
\end{equation}
The time evolution of $v_s$ and $\T$ is given by:
\begin{equation}
    \begin{aligned}
    v_s(t) & = 0.1c\cdot\left(\frac{1}{f_{\Omega}}\frac{\left(1-k\right)}{\left(s-1\right)}\right)^{\frac{1}{3}}L_{bo,45}^{\frac{1}{3}}t_{bo,d}^{k-\frac{1}{3}}t_d^{-k}, \label{eq:vs_of_t}
\end{aligned}
\end{equation}
\begin{equation}
\begin{aligned}
        \T(t)  &=10\left(\frac{1}{f_{\Omega}}\frac{\left(1-k\right)}{\left(s-1\right)}\right)^{-\frac{1}{3}} L_{bo,45}^{-\frac{1}{3}}t_{bo,d}^{b+\frac{1}{3}}t_d^{-b}.
        \end{aligned}
\label{eq:tau_of_t}
\end{equation}
where $t_{bo,d}=t_{bo}/1$d and $t_d=t/1$d. From here on we neglect $\left(\frac{1}{f_{\Omega}}\frac{\left(1-k\right)}{\left(s-1\right)}\right)$, as it is of order unity. 
We start by finding the break times $t(IC\to ff),t(\T=1),t(v_9\simeq0.3), t(v\simeq0.1)$. These times define three main regimes: (i) fast IC, for which, as seen in Figure \ref{fig:phase_space}, $\T$ is never much smaller than unity, so we can always approximate $\T\gtrsim 1$ in this regime. (ii) fast free-free with $\T>1$; (iii) fast free-free with $\T<1$, Both free-free regimes are further divided into three sub-regimes, by the shock velocity, where for $v_s\ge 3\cdot 10^8{\rm ~cm/s}$ there is emission above 10 keV, $3\cdot 10^8{\rm ~cm/s}\ge v_s\ge 1\cdot 10^8{\rm ~cm/s}$ means there is X-ray emission only below 10 keV, and for $v_s\le 1\cdot 10^8{\rm ~cm/s}$, the shocked matter is cooled primarily by line cooling, and no thermal X-rays are expected. Once the shock velocity approaches $\sim 3\cdot 10^8 {\rm ~cm/s}$, the expressions for the X-ray light curves are not power-laws, since the spectral cut-off approaches 10 keV. As a result, the expression becomes inconvenient to write in the manner expressed in this section. We therefore only derive the times of these breaks, and limit our explicit light-curves to $v\ge 4\cdot 10^8{\rm~cm/s}$. For lower velocities, the light curve can be derived from the previous chapter.
We summarize the values of the break time in Table \ref{tab: break_times}. 
\begin{table*}
\centering
\caption{Characteristic light-curve break times in terms of $(\tau_{bo},t_{bo})$ with $\tau_{bo}= c/v_{bo}$, and in terms of $(t_{bo,d},L_{bo,45})$}
\label{tab: break_times}
\begin{tabular}{lccc}
\hline
\hline
Break & $t$ in terms of $\tau_{bo},t_{bo}$ & $t$ in terms of $t_{bo,d},L_{bo,45}$ & Significance\\
\hline
$\T=1$ 
& $t_{bo}\,\tau_{bo}^{1/b}$ 
& $1\,\mathrm{d}\cdot 10^{1/b}\!L_{bo,45}^{-1/(3b)}\,t_{bo,d}^{\,1+1/(3b)}$
& Hard X-ray hump joins the soft continuum\\[8pt]

$IC\!\to\! ff\;(\T>1)$
& $t_{bo}\,35^{\frac{1}{k+\frac{b}{4}}}\,\tau_{bo}^{-\frac{3}{4k+b}}$
& $1\,\mathrm{d}\cdot 1500^{\frac{1}{4k+b}}\!
L_{bo,45}^{\frac{1}{4k+b}}\;
t_{bo,d}^{\,1-\frac{1}{4k+b}}$
& $L_{\rm opt}\simeq L_{x,{\rm soft}}/\max(\T,1)$, shock instability onset \\[8pt]

$v_s=3\cdot10^{8}\,\mathrm{cm\,s^{-1}}$ 
& $t_{bo}\,100^{1/k}\,\tau_{bo}^{-1/k}$ 
& $1\,\mathrm{d}\cdot 10^{1/k}\!L_{bo,45}^{1/(3k)}\,t_{bo,d}^{\,1-1/(3k)}$
& Hard X-rays disappear\\[8pt]

$v_s=10^{8}\,\mathrm{cm\,s^{-1}}$ 
& $t_{bo}\,300^{1/k}\,\tau_{bo}^{-1/k}$ 
& $1\,\mathrm{d}\cdot 30^{1/k}\!L_{bo,45}^{1/(3k)}\,t_{bo,d}^{\,1-1/(3k)}$
& X-ray continuum disappears\\
\hline
\end{tabular}
\end{table*}

With these, we can now evaluate the ratios from Table \ref{tab:ratios_expr}, and express them too in terms of measurable parameters, in Table \ref{tab:ratios_t}, and then adapt the equations from the previous section and summarize the light curve in Table \ref{tab:lightcurve}. 
\begin{table*}
\centering
\caption{Same as Table \ref{tab:ratios_expr}, but in terms of the measurable parameters and time}
\begin{tabular}{|c|c|c|}
\hline
Regime & $L_{x,\mathrm{soft}}/L_{\mathrm{opt}}$ & $L_{x,\mathrm{hard}}/L_{x,\mathrm{soft}}$ \\
\hline\hline

fast IC ($\T \gtrsim 1$) 
& 
$10^{-4}\tau_{bo}\left(\frac{t}{t_{bo}}\right)^{\frac{4b+7k}{3}}
\approx 10^{-3}L_{bo,45}^{-\frac{1}{3}}t_{bo,d}^{\frac{1-4b-7k}{3}}t_d^{\frac{4b+7k}{3}}$
&
$0.8\tau_{bo}\left(\frac{t}{t_{bo}}\right)^{\frac{k-2b}{3}}
\approx 8\cdot L_{bo,45}^{-\frac{1}{3}}t_{bo,d}^{\frac{1-k+2b}{3}}t_d^{\frac{k-2b}{3}}$
\\ \hline

fast free-free, $\T>1$ 
& 
$\tau_{bo}^{-1}\left(\frac{t}{t_{bo}}\right)^b
\approx 0.1L_{bo,45}^{-\frac{1}{3}}t_{bo,d}^{\frac{1}{3}-b}t_d^{b}$
&
$95\left(\frac{t}{t_{bo}}\right)^{-b-k}
\approx 95\cdot t_{bo,d}^{b+k}t_d^{-b-k}$
\\ \hline

fast free-free, $\tau_T\lesssim1$
&
$1$
&
$95\left(\frac{t}{t_{bo}}\right)^{-k}
\approx 95\cdot t_{bo,d}^{k}t_d^{-k}$
\\ \hline

\end{tabular}
\label{tab:ratios_t}
\end{table*}

\begin{table*}
\centering
\caption{The light curve for the three regimes of interest (here $k_B=1$).}
\label{tab:lightcurve}
\tiny
\begin{tabular}{|c|c|c|c|}
\hline
Regime & $L_{opt}$ & $L_{x,soft}$ & $L_{x,hard}$\\
\hline\hline

fast IC &
$L_{s}=L_{bo}\!\left(\frac{t}{t_{bo}}\right)^{1-b-4k}$ &
$\frac{L_s}{\T}\left(\frac{10{\rm ~keV}}{\Ts}\frac{T_{ff=IC}}{\Ts}\right)^{1/2}=10^{-4}L_{bo}\tau_{bo}\left(\frac{t}{t_{bo}}\right)^{\frac{3+b-5k}{3}}$ &
$L_{s}\frac{T_{ff=IC}}{T_{s}}=10^{-4}L_{bo}\tau_{bo}^{2}\left(\frac{t}{t_{bo}}\right)^{\frac{3-b-4k}{3}}$ \\
($\tau\gtrsim1$)&&&\\
 &
$\approx10^{45}\frac{\mathrm{erg}}{\mathrm{s}}\,L_{bo,45}\,t_{bo,d}^{-1+b+4k}t_d^{1-b-4k}$ &
$\approx10^{42}{\rm erg/s}\,L_{bo,45}^{2/3}t_{bo,d}^{\frac{-4-b+5k}{3}}t_d^{\frac{3+b-5k}{3}}$ &
$\approx10^{43}\frac{\mathrm{erg}}{\mathrm{s}}\,L_{bo,45}^{1/3}t_{bo,d}^{-\frac{1-b-4k}{3}}t_d^{\frac{3-b-4k}{3}}$ \\
\hline

fast free-free &
$L_{s}\!\left(\frac{10\,\mathrm{keV}}{T_{s}}\right)^{1/2}=10^{-2}L_{bo}\!\left(\frac{t}{t_{bo}}\right)^{1-b-3k}$ &
$\frac{1}{\tau}L_{s}\!\left(\frac{10\,\mathrm{keV}}{T_{s}}\right)^{1/2}=10^{-2}L_{bo}\!\left(\frac{t}{t_{bo}}\right)^{1-3k}$ &
$L_{s}=L_{bo}\!\left(\frac{t}{t_{bo}}\right)^{1-b-4k}$ \\
($\tau_T>1$) &&&\\
&$\approx10^{43}\frac{\mathrm{erg}}{\mathrm{s}}\,L_{bo,45}t_{bo,d}^{-1+b+3k}t_d^{1-b-3k}$ &
$\approx10^{43}\frac{\mathrm{erg}}{\mathrm{s}}\,L_{bo,45}t_{bo,d}^{-1+3k}t_d^{1-3k}$ &
$\approx10^{45}\frac{\mathrm{erg}}{\mathrm{s}}\,L_{bo,45}t_{bo,d}^{-1+b+3k}t_d^{1-b-3k}$ \\
\hline

fast free-free &
\multicolumn{2}{c|}{$\frac{1}{2}L_{s}\!\left(\frac{10\,\mathrm{keV}}{T_{s}}\right)^{1/2}=5\cdot10^{-3}\tau_{bo}L_{bo}\!\left(\frac{t}{t_{bo}}\right)^{1-b-3k}$} &
$L_{s}=L_{bo}\!\left(\frac{t}{t_{bo}}\right)^{1-b-4k}$ \\
($\tau_T\lesssim1$) &\multicolumn{2}{c|}{}&{}\\
&\multicolumn{2}{c|}{$\approx5\cdot10^{43}\frac{\mathrm{erg}}{\mathrm{s}}\,L_{bo,45}^{2/3}t_{bo,d}^{-\frac{2}{3}+b+3k}t_d^{1-b-3k}$} &
$\approx10^{45}\frac{\mathrm{erg}}{\mathrm{s}}\,L_{bo,45}t_{bo,d}^{-1+b+3k}t_d^{1-b-3k}$ \\
\hline

\end{tabular}
\end{table*}

\section{Global radiative shock instability}\label{sec:instability}
Radiative shock waves can undergo global thermal instabilities (technically, an overstability), depending on the cooling processes in the post-shock region \citep{Langer1981}.
Using a linear analysis, \cite{Chevalier1982C} show that a planar radiative shock is unstable to the first and second overtone modes when free-free emission
dominates the cooling, i.e., $\Lambda\propto T^{1/2}$, 
while it is marginally stable in the IC cooling regime, i.e., $\Lambda\propto T$. Here, $\Lambda$ denotes the cooling function (energy times volume per unit time). 
In the free-free cooling regime, the shock front oscillates with periods approximately the cooling time of the post-shock region: $P_{\rm osc}\approx 1.8t_{\rm ff}$. The amplitude of the oscillation grows exponentially with an e-folding of several oscillation periods.

With one-dimensional hydrodynamics simulations, \cite{Imamura1984ApJ,Migonne2005} confirm the instability condition derived from the linear analysis. 
The amplitude of the shock-front oscillations saturates in 1D at roughly the $20\%$ level, corresponding to the luminosity variations of approximately a factor of two.
In two-dimensional simulations, \cite{Sutherland2003ApJ} find that
the inhomogeneous structure of the postshock density enhances cooling (most likely due to mixing), resulting in somewhat shorter oscillation periods than those expected from the linear analysis. 

In our model, the onset of the instability is expected to occur around the transition from the IC to free-free fast cooling regimes (equation \ref{eq:v_tr}). The oscillation period of the first overtone mode is
\begin{equation}\label{eq:Posc}
\begin{aligned}
 P_{\rm osc}& \approx
    1\,{\rm day}\,\left(\frac{v_{bo}}{0.1~c}\right)^{3} b^{-1}  t_{bo,d}^{k (s+1)-s+1}t_{d}^{s-k(1+s)}\\
    &\approx 1.7 \,{\rm day}\,\left(\frac{v_{bo}}{0.1~c}\right)^{3} t_{bo,d}^{-0.6}t_{d}^{0.4}.
 \end{aligned}  
\end{equation}
where the second equality is for $k=0.6$ and $s=2.5$. Note that global oscillations are expected when the cooling time is not much shorter than the dynamical time. In the case of $t_{\rm ff} \ll \tdyn$, the thin-shell instability is expected to occur (\cite{Vishniac1994}, see also simulations by \citep{Steinberg2018}), and break the shock fronts into small sections that oscillate individually, rather than produce global oscillations. This limits the region of phase space where we expect the global radiative instability to occur, to where the cooling time should be smaller than the dynamical time, but not by too much. Moreover, for the global oscillations to be observable, the system must remain in a marginally slow cooling state for several dynamical times, rather than quickly transitioning between fast and slow cooling.
Even when $\tff\sim \tdyn$, it is possible that the oscillation will develop in multiple distinct regions, reducing the observed oscillation amplitude. We estimate distinct regions should be no smaller than the width of the cooling length, which for a marginally adiabatic shock is $\frac{\Delta r}{r}\sim \frac{1}{10}$, thus we suspect that there should be no more than $\sim 10$ distinct oscillating regions, and possibly fewer. This effect, too, should be investigated in 3D simulations. 

\section{interpretation of multi-wavelength observations of AT2018cow}\label{sec:2018cow}
\subsection{X-ray and optical light-curves}
Applying the model from \S\ref{sec:optical_X-ray} to AT2018cow, we find that a significant part of the parameter space identified there, i.e., $v_{bo}\approx 0.1$ c, $t_{bo}\approx 1$ d, $s\simeq 2.4-3.1$, $k\simeq \frac{4-s}{5-s}$, can also explain all the other properties of the x-ray and optical/UV emission\footnote{The model in \S\ref{sec:optical_X-ray}  assumes that X-rays are not absorbed in the shock upstream. To verify consistency, we use the formalism of \cite{GS2025} and find that, indeed, no absorption is expected. This fits the constraints from observations, which find that the neutral equivalent column density is consistent with galactic origin \citep{Margutti2019,Sandoval2018}}. This includes the luminosity normalization, the time evolution of the ratios between the different bands, and the rapid soft X-ray variability. An example of a good fit is  $t_{bo}= 0.7 {\rm ~d}$, $v_{bo}=0.13c$, $s=2.5$, and $k=0.6$. This corresponds to $r_{bo} = 6\cdot 10^{14} {\rm ~cm}$ and to a CSM density profile, and shock radius and velocity evolution of:
\begin{equation}\label{eq:rho}
    \rho = 4.5\cdot 10^{-14}{\rm ~gr/cm^3} \left(\frac{r}{r_{bo}}\right)^{-2.5},
\end{equation}

\begin{equation}\label{eq:vs}
    v_s \approx 30,000 {\rm~km~s^{-1}} ~t_d^{-0.6} ~~~;~~~ 1d\lesssim t <40 d~,
\end{equation}
and 
\begin{equation}\label{eq:rs}
    r_s \approx 6 \times 10^{14} {\rm~cm} ~t_d^{0.4} ~~~;~~~ 1d\lesssim t <40 d~,
\end{equation}
so the mass collected by the shock at time $t$ is roughly 
\begin{equation}\label{eq:M}
    M(<r_s) \approx 0.15 {\rm~M_\odot} ~t_d^{0.2} ~~~;~~~ 1d\lesssim t <40 d  ~.
\end{equation}

\begin{figure}
    \centering
\includegraphics[width=\linewidth]{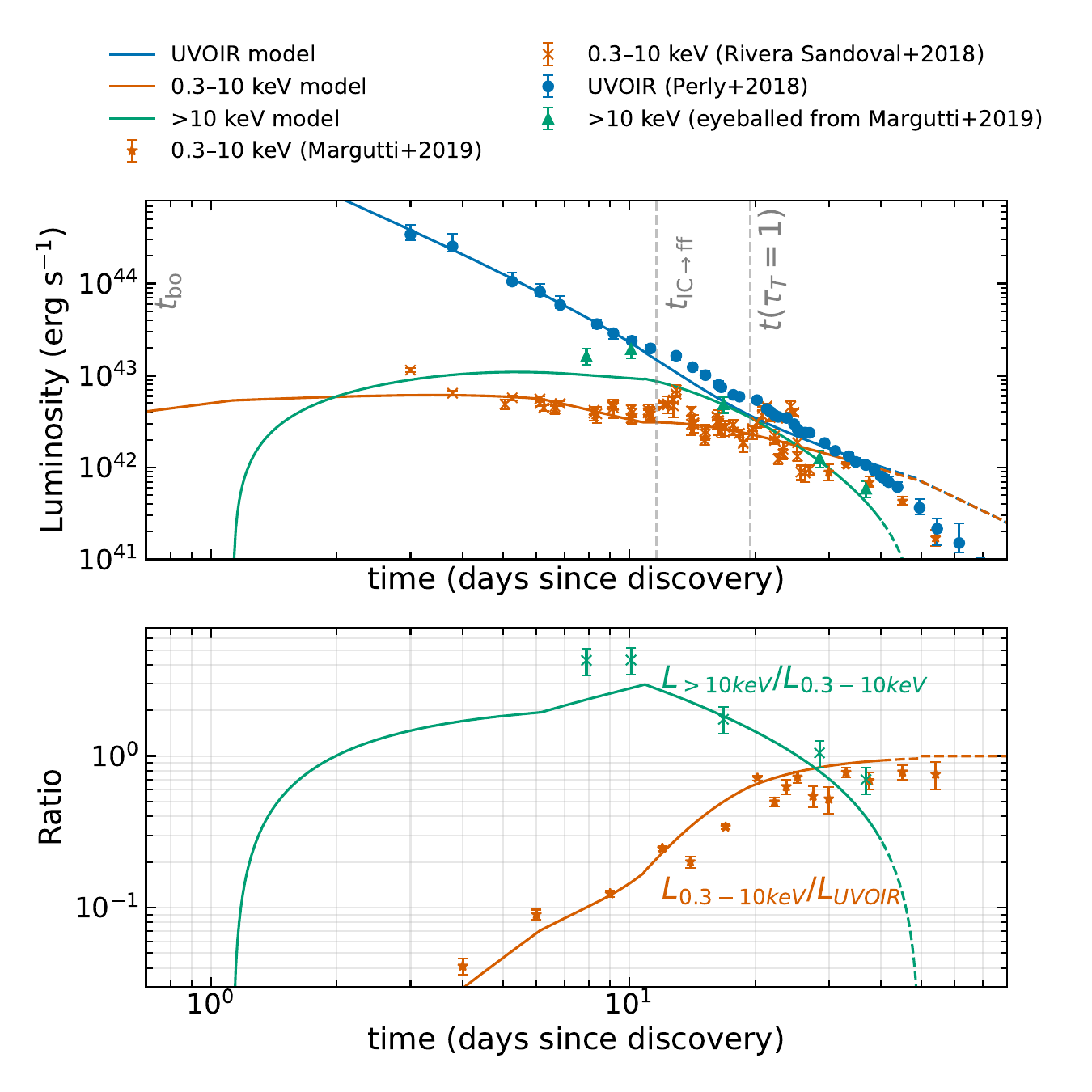}
    \caption{A comparison of our model with $t_{bo}= 0.7 {\rm ~d}$, $v_{bo}=0.13c$, $s=2.5$, and $k=0.6$, to the optical and X-ray observations. The top panel shows light curves, and the bottom panel shows the ratios between the different bands.}
    \label{fig:lightcurve}
\end{figure}

In Figure \ref{fig:lightcurve} we plot the observed light curves and the ratios between the different bands alongside our model with these parameters. The model in the figure accounts for various effects of order unity, neglected in the model above, such as the emission from the sub-dominant cooling process (i.e., IC emission in the free-free dominated regime, and vice versa). These add corrections of less than a factor of 2, and result in a smooth light curve. We also mark in the figure the shock breakout time, and the transitions from IC to free-free cooling and from Thomson thick to Thomson thin. These plots exemplify the visual effect the transitions between different regimes have on the light curve. 

The top panel shows a good fit between the observations and the model till day $\sim 40$, after which the observations decline more steeply. The bottom panel shows that the ratio of soft X-ray to optical luminosity remains as predicted, also at later times. Thus, we conclude that the hydrodynamic model of a single shock in thick CSM, which is fully constrained by four parameters, reasonably fits the observations till day 40. We infer that around day 40, something changes that either reduces the total luminosity (i.e., the shock reaches the end of the dense CSM, or a place where the CSM solid angle decreases sharply), or shifts growing fractions of the luminosity to the unobserved $10 {\rm~eV}-0.2{\rm~keV}$ range, though the latter is disfavored by the X-ray observation at 220 days (see \S\ref{sec: Late CSM}). 
Note that between day 40 and day $\sim$60, the soft X-ray/Optical luminosity ratio implies that despite the changes to the hydrodynamics, the reprocessing mechanism keeps operating as expected.  

The soft X-ray spectrum is as expected; $L_\nu \sim \nu ^{-1/2}$, and at around day 8, shows an additional feature \citep{Margutti2019}, possibly blue-shifted iron lines, further discussed in \ref{sec:lines}
\subsection{X-ray fluctuations}
We expect the radiative shock instability discussed in \S\ref{sec:instability} to start taking place at the transition from fast IC to fast free-free cooling, around day 10. The X-ray fluctuation time scale increases from about 4 days at day 10 to about 10 days at day 60 \citep{Margutti2019,Kuin2019}. 
Remarkably, the oscillation periods of the instability are comparable to the observed oscillations.  The periods predicted by the 1D model with our canonical parameters (Eq. \ref{eq:Posc})  are about twice the observed periods, and in 3D we expect it to be smaller by some factor (e.g., in the simulations of \citealt{Sutherland2003ApJ} the periods in 2D are shorter by a factor of about 1.35 compared to 1D).  
We therefore conclude that radiative shock instability predicts X-ray fluctuations that agree to within a factor of order unity with the amplitude and time scales of the observed fluctuations, including the increase of the timescale by a factor of $\sim 2$ between day 10 and day 60. Further research on this instability in 3D is needed to make a more accurate quantitative comparison with observations. Issues that are needed to be resolved, in addition to more accurate estimates of the period and amplitude, are: (i) what the coherent scale of the global instability is (i.e., how many independent patches, if any, are expected along the shock front), and (ii) in which regime the thin shell instability takes over and breaks the shock front in multiple places.

As the optical photons are reprocessed soft X-rays, the shock oscillations can affect them as well. However, several effects are expected to make the optical light curve much smoother than the X-ray light curve. First, during reprocessing, photons can be absorbed and re-emitted several times before finally radiated as optical photons towards the observer. Second, the entire system is marginally optically thin to optical photons but highly optically thick to X-rays. Thus, while the X-ray we observe are emitted from the shock regions along our line of sight, the optical photons that we see are either emitted directly or back-scattered toward us from the far side of the system, arriving with a delay of twice the light crossing time of the entire system compared to the X-ray photons that were emitted alongside them. The light crossing time of the system (from side to side) is:
\begin{equation}
    t_{lt} \approx \frac{2v_{bo}}{(1-k)c} t_{bo}^{k} t^{1-k}= 1.7 {\rm~d~} \left(\frac{v_{bo}}{0.1c}\right) t_{bo,d}^{0.6} \left(\frac{t}{20{\rm ~d}}\right)^{0.4}
\end{equation}
Thus, $2t_{lt}$ is comparable to the fluctuation time scale. Finally, the difference in the optical depth of the system to optical and X-rays causes an additional smoothing effect, as it implies that in optical, we see multiple regions that are out of phase. Thus, our model predicts optical fluctuations that are much weaker than those seen in X-rays.         
Examination of the multi-wavelength optical light curve (Fig. 2 in \citealt{Perley2019}) shows a multi-wavelength excess that lasts a few days around day 20 and day 29, which coincide with X-ray flares. The optical variability is much weaker than the variability seen in X-rays, and it is not seen in coincidence with all observed X-ray flares. We therefore conclude that our model is broadly consistent with the observed variability in the various bands, although detailed quantitative modeling is required to verify or rule out this expectation.
If indeed this is the explanation for the X-ray variability, the limited range of parameters for which a prolonged variability phase is expected means that we do not necessarily expect to observe such variability in other interaction-powered transients, or even in LFBOTs with slightly different parameters. 

\subsection{The blackbody radius and IR emission}\label{sec:optIR}
Our model predicts that the optical and IR originate from the reprocessed X-ray emission. This emission is from approximately the shock radius, which is much larger than the blackbody radius inferred from observations. To check whether the observed emission is consistent with reprocessed emission, we ran a CLOUDY \citep{CLOUDY2023} simulation for the conditions on day 30, when the matter becomes sufficiently Thomson-thin for such a simulation to be marginally valid.  We obtain an approximation of the reprocessed optical emission from the cold downstream by simulating the reflection of free-free X-ray emission that illuminates a constant-pressure region. We set the pressure and column density of this region to emulate the downstream conditions in our model on day 30, and assume solar metallicity. This calculation is not fully accurate, as it includes only the coldest gas in the system, with $T \approx 10^4$ K, whose conditions are set by an equilibrium between the absorbed X-rays and the emitted optical radiation. This model does not include the cooling and the reflected emission from the gas that is cooling from $\sim 1$ keV to the temperature of the illuminated slab in the simulation. Thus, we expect this model to provide a reasonable approximation of the optical and IR emission, and to fall short in the UV. We also expect the combined emission from gas at different temperatures to smooth the spectrum significantly compared to CLOUDY's result.
We stress that the parameters used in this CLOUDY simulation are taken from the model shown in Fig. \ref{fig:lightcurve}, which is based on X-ray and optical light curves only, with no additional free parameters and with no information on the exact spectrum predicted by this model. 

\begin{figure}
    \centering    \includegraphics[width=\linewidth]{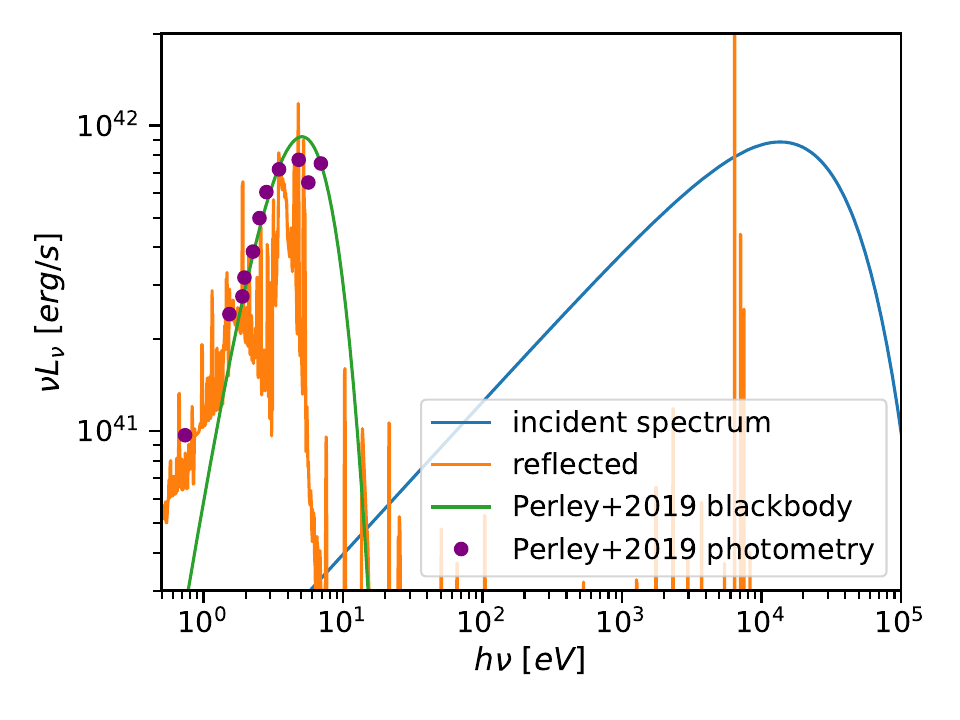}
    \caption{The predicted optical-IR spectrum from a CLOUDY simulation of the X-ray spectrum at day 30 reflected off a cold dense shell. The reflected spectrum is smoothed on a frequency scale corresponding to the expected Doppler broadening, though it is likely that additional broadening factors are also present.  
    We plot the observations and model of \citep{Perley2019} for reference. 
    Note that this simulation is a crude approximation that does not include all the physics; specifically, it does not include the regions at UV temperatures or their emission.}
    \label{fig:cloudy}
\end{figure}

Fig. \ref{fig:cloudy} shows the incident X-ray spectrum, and the reflected spectrum (smoothed, on the scale expected for Doppler broadening) from a constant-pressure slab at the shock radius as calculated by CLOUDY, alongside the observations on day 30 reported in \cite{Perley2019}, and the blackbody portion of the fit they use. 

This figure shows, first, that our model explains the observed spectrum reasonably well in the optical, including the IR deviation from blackbody, and lacks energy in the UV, as expected. The IR excess is most likely due to the contribution of free-free emission, since the spectrum of the continuum below the peak seems to be roughly flat in $F_\nu$ (\cite{Chen2025} reached a similar conclusion as to the origin of the IR excess and deviation from a black-body in a reprocessed spectrum, though the geometry they assume differs significantly from ours.). Second, this figure shows
that although the emission originates at the radius of the shock on day 30, $\sim 3 \times 10^{15}$ cm, the observed spectrum and luminosity result in a much smaller blackbody radius  $R_{BB}=\left(\frac{L}{4\pi\sigma_{B} T^4}\right)^{1/2} \sim 2 \times 10^{14}$ cm, when the temperature is estimated based on the spectrum near the peak of $\nu F_\nu$ ($\sigma_{B}$ is the Stefan-Boltzmann coefficient). The reason is that the radiation reflected from the cooled gas in the downstream of the shock is far from thermodynamic equilibrium, and although its spectrum resembles a blackbody with a temperature of the gas in the reflecting region $T$, its energy density at the source is far below $a_{BB} T^4$, where $a_{BB}$ is the radiative constant.
As the luminosity drops with time, the decrease in $\T$ reduces the number of scatters behind the shock, and the radiation falls further from thermodynamic equilibrium. 
In this out-of-thermodynamic-equilibrium state, the temperature is set by a balance of heating and ionization by the X-rays, and cooling by line emission. It remains in the range of $1-2 \times 10^4$ K, as the cooling function at these energies depends very strongly on the temperature in this range, acting as a thermostat. Thus, the spectral shape of the emission reflects the blackbody shape of the roughly constant gas temperature. Since the system is moving farther from thermodynamic equilibrium, the deviation of the observed spectrum from a blackbody increases with time. 

The nearly constant temperature and decreasing luminosity cause the measured blackbody radius, $R_{BB}$, which assumes that the source is at thermodynamic equilibrium, to drop with time, while in reality the radius at which the radiation originates increases monotonically. To estimate the evolution of $R_{BB}$, we can use the fact that the temperature of the illuminated gas does not change much during the optically thin phase,  but as its luminosity decreases, the blackbody radius decreases as well, roughly as $L^{1/2}$ after day 20. This example shows that one should be cautious when using $R_{BB}$ as an estimate of the emitting region's location when the source is out of thermodynamic equilibrium.

\subsection{Radio and Sub-millimeter}
In our model, the dense CSM must cover a significant part of the $4\pi$ sr, but it must also have a non-negligible part where the CSM has a much lower density. Since the ejecta is unlikely to be confined only to the direction of the dense CSM, our model predicts another potential source of emission due to the interaction of the ejecta with the low-density CSM. Below, we verify that the observed radio emission is consistent with being the result of this interaction, and see what the resulting constraints on the ejecta are. 

Radio observations between days 22 and 132 \citep{Margutti2019,Ho2019} suggest shock-powered synchrotron emission, with a spectrum that peaks at the synchrotron self-absorption frequency. According to this interpretation the shock has a roughly constant velocity of about $0.1$ c propagates in a CSM with a density of about $10^{-18} {\rm g~ cm^{-3}}$  (to within an order of magnitude) at a radius  $r \sim 6 \times 10^{15}$ cm and a density of about  $4\cdot 10^{-20} {\rm g~ cm^{-3}}$ at $r \sim 3 \times 10^{16}$ cm (these values depend on the exact assumptions; \citealt{Margutti2019}). 
The roughly constant shock velocity implies that it is driven by ejecta with a similar, or at most slightly higher, velocity and with mass that is larger than the mass it collects until day 132, $\sim 10^{-2} M_\odot$. This is fully consistent with our model, as long as the ejecta velocity is $v_{ej} \approx 0.1$ c, in which case its mass is $m_{ej} \sim 0.01- 0.05~M_\odot$ (see \S\ref{sec:ejecta}), high enough to be consistent with the radio observations.

The sub-mm emission seems to be generated by the same source as the radio emission, namely by a synchrotron emission from the shock that propagates into the dilute part of the CSM \citep{Ho2019,Margutti2019}. At early times, the optically thin spectra observed in sub-mm wavelengths were significantly steeper than the usual spectra observed in synchrotron sources \citep{Ho2019}. This was suggested to result from the contribution of electrons near the thermal peak \citep{Margalit2021,Margalit2024}.

Note that all the radio and sub-mm observations in which a SSA peak is observed are from radii larger than the radius of the dense CSM probed by the X-ray emitting shock at day $\sim 40$. At day 40, the X-ray emitting shock is at $r\sim 3\cdot 10^{15}\rm~ cm$ and probes a density of $\simeq 8\cdot 10^{-16} {\rm gr/cm^3}$. Comparing this to the density probed by the radio-emitting shock extrapolated to the same radius (assuming an $\rho\propto r^{-2}$ density profile), we find the dense CSM is more than two orders of magnitude denser than the dilute CSM, at a few times $10^{15} {\rm~cm}$.

\subsection{Spectral lines}\label{sec:lines}
Our model predicts broad spectral lines that form in the downstream of the shock, and narrow lines from the unshocked upstream, as is commonly seen in interaction-powered transients. Broad lines are expected to have a width corresponding to a velocity of $\sim v_s$, while narrow lines appear only once the upstream temperature is low enough for them to form (at high temperatures, ions are significantly less likely to recombine, and line emission is suppressed), and when the matter is Thomson thin, so that the lines are not broadened by scattering with hot electrons. The upstream becomes Thomson thin around day 20 and calculating the temperature of the upstream based on the X-ray emission, as described in \citep{GS2025}, we find that the matter is at the Compton temperature, i.e., $\sim 10 \rm ~keV$ until sometime between day 10 and 15, and starts decreasing afterwards, reaching about $\sim 0.1 \rm ~keV$ at day 30. We therefore expected narrow lines to start appearing only around day 20 or even later, and for their strength to increase with time.  See also \cite{Aspegren2026} for a study of line suppression, though it is unclear which of their results apply to a reflected spectrum, and their use of the LTE approximation is likely not valid for our system, in which there are high ionizing radiation fluxes in the regions forming the optical/
UV spectra.
We find that the observed spectral features are broadly consistent with this prediction; the width of broad lines roughly corresponds to the shock velocity, while the narrow lines appear at around day $22$, once the matter is no longer Thomson thick, and the upstream temperature is decreased.

An additional noteworthy feature is the spectral emission detected around 8 keV in the 7.7-day X-ray observation and its subsequent disappearance by day 15,  attributed to iron lines \citep{Margutti2019}. A comprehensive investigation of iron lines cannot be performed using CLOUDY and is beyond the scope of this work, so here we offer speculation regarding the origin of this feature and its disappearance based on our model framework.
We expect iron line emission from both the upstream and downstream regions, with the most prominent lines arising from the $n=2\to 1$ transition. These include the 6.4 keV fluorescence line from weakly ionized iron, the 6.7 keV line from He-like iron at temperatures of a few to 10 keV, and the 7 keV line from H-like iron at temperatures around 10 keV. The upstream region, should emit the 6.7 and/or 7 keV line while the plasma remains above the Compton temperature (until day 10-15), and than cease one the temperature drops below a few keV. We speculate that the emission feature at day 7.7 originates from the 6.7 or 7 keV line, which is scattered in the upstream, which is at a temperature of $\sim 10 \rm~ keV$. Since the Thomson optical depth of the upstream at this time is $\T\sim 1.5-2$, photons undergo 3-4 Compton scatters which blue-shift the lines to $8-9$ keV, and broaden them. This interpretation naturally explains the feature's disappearance: once the upstream cools, it no longer emits these lines. 
To fully model the line evolution, the exact iron ionization state and temperature need to be modeled, and the contributions from both the upstream and the downstream must be considered.

\subsection{Late time blue emission}
The late-time UV-optical HST observations and X-ray observations/upper limits provide a lower limit of $\sim 10^{39} {\rm erg/s}$ on the bolometric luminosity between day $\sim 700$ and $\sim 2000$ days, which seems to be slightly decreasing with time. The UV-optical observations are consistent with the Rayleigh-Jeans tail of a source with $T_{BB}\approx3\cdot 10^4-6\cdot 10^5 K,L_{BB}\simeq 2\cdot 10^{39}-5\cdot 10^{43}{\rm erg/s}$ at day 700  and $T_{BB}\simeq 2\cdot 10^4-2\cdot 10^5 {\rm K}, L_{BB}\simeq 2\cdot 10^{39}-3\cdot 10^{41}{\rm ~erg/s}$, where we report here the unified estimates of \cite{Chen2023} and  \cite{Inkenhaag2023}. If the source of this emission is in thermodynamic equilibrium, then its radius is $\sim 5 \times 10^{12}$ cm. X-ray observations with XMM-Newton find a luminosity of $\simeq2\cdot 10^{39} {\rm erg/s}$ at 218 days, and of $\sim 4\cdot 10^{38} {\rm ~erg/s}$  at $\sim$1350 days \citep{Migliori2024}. The observed emission at 218 days is clearly from AT2018cow. At day 1350, it is unclear whether it is due to the source or the underlying host galaxy. In both epochs, the spectrum is soft, and photons are only observed up to $\sim 4.5 {\rm ~ keV}$. 

Our model does not provide any predictions for these late times, except that the shock that propagated into the dense CSM continues to propagate and may or may not produce observable emission, depending on the CSM density profile it probes between day 40 and these later epochs. The shock at these late epochs may be slower, or maintain a similar velocity to that observed in our last modeled epoch (at $\sim 40$ days), depending on whether it collected additional mass, or if it reached the edge of the dense CSM and started coasting.  Here, we briefly discuss the interaction of this shock with the CSM as a potential source of late emission, as well as two additional potential sources for this emission: a remaining companion star and an obscured additional energy source, such as an accretion disk. As our model is not constraining for this stage, all three sources of the emission, and maybe others we did not think of, can be compatible with our model.

\subsubsection{Late-time emission from CSM interaction}\label{sec: Late CSM}
The most natural source of the emission within the framework of our model would be if the shock reached the edge of the dense CSM at around day 40, causing the observed decline in the luminosity, and then coasted through the surrounding dilute medium. The radio-emitting shock probed the dilute medium between a radius of $\sim 5\cdot 10^{15} {\rm ~cm}$ and $\sim 3\cdot 10^{16} {\rm ~cm}$. In this range, \cite{Margutti2019} find that assuming a shock velocity of $0.1$c and that the magnetic field takes a fraction of 0.01-0.1 of the internal energy behind the shock, the dilute medium density profile is $\rho = \rho_0 \left(\frac{r}{10^{16}{\rm~cm}}\right)^{-2}$ where  
\begin{equation}
    \rho_0 \sim  0.5-5 \times 10^{-19} {\rm~gr~cm^{-3}}, ~~~;~~~ 0.5 \lesssim r_{16} \lesssim 3
\end{equation}
In our model, the radius of the shock that produces the early optical and X-ray emission is about $3\cdot 10^{15}~\rm cm$ at day 40, and its velocity is $\sim 3 \cdot 10^8 {\rm~cm~s^{-1}}$. Given the uncertainty in the shock velocity at day 40 and its evolution until it starts coasting in the dilute CSM, we consider a shock velocity in the range  $ 1-5 \cdot 10^8 {\rm~cm~s^{-1}}$ 

At these velocities and at the density of the dilute CSM the shock is slow cooling ($\tff, t_{IC}>\tdyn$, see equation \eqref{eq:ff_fast_condition}). The shocked gas emits a fraction of its energy in X-rays, while cooling adiabatically\footnote{In this analysis, we consider only contributions from the forward shock. Depending on the density profile, the reverse shock may, or may not, produce significant emission comparable to that of the forward shock. If it does, it may change the parameters needed to explain the late time emission.}: 
\begin{equation}\label{eq: x slow}
\begin{aligned}
        L_{x,slow}&\simeq L_s\frac{\tdyn}{\tff}=2\cdot 10^{38} {\rm ~erg/s~} 
 \rho_{-18}^2 t_{1000}^3 v_8^4\\&= 4\cdot 10^{36} {\rm ~erg/s~}  \rho_{0,-19}^2 t_{1000}^{-1},
 \end{aligned}
\end{equation}
where $t=t_{1000} \cdot 1000$d.
As the matter in the downstream cools adiabatically, it eventually reaches the temperature at which line cooling takes over and becomes fast enough to radiate the remaining energy within one dynamical time, in UV/optical. 
Using the approximate line cooling function $\Lambda\simeq 1.1\cdot 10^{-22} {\rm ~erg\cdot cm^3/s~T_6^{-0.7}} ~~~;~~~ T\lesssim 3\cdot 10^7 \rm 
~K,$ \citep{Draine2011}, the line cooling timescale is: 
\begin{equation}
    t_{lines}=\frac{3}{2}\frac{m_pk_BT_e}{\rho\Lambda}\simeq 36 {\rm ~d~} T_{6}^{1.7}\rho_{-18}^{-1} 
\end{equation}
Assuming it is in the coasting phase, the reverse shock keeps the post-shock pressure constant. We can solve for $t_{lines}=\tdyn$ at constant pressure set by the shock jump conditions, and find that line cooling becomes dominant over adiabatic losses at: 
\begin{equation}
    T_{lines} = 1.4\cdot 10^7 {\rm ~K~}\rho_{-18}^{0.4} t_{d,1000}^{0.4} v_8^{0.8}. 
\end{equation}
The luminosity from the line cooling region can be calculated by finding the internal energy remaining in the line-cooling gas after the adiabatic losses, and dividing it by the current dynamical time (which is equal to the line-cooling time of this region). When this matter passed the shock, it was given an energy $L_s't'$, where we use $'$ to denote the retarded time and the properties at this time. Since the shock velocity is constant and $\rho \propto r^{-2}$, the shock luminosity is approximately constant, as is the temperature in the immediate downstream of the shock. 

Consider a gas element that crosses the shock at $t'$, and is pressurized to $P'$, and given a temperature $T'=\Ts$ where $\Ts$ is the temperature immediately behind the shock which does not change with time. The element cools adiabatically up to time $t$, when its pressure is $P$, and its temperature is $T_{line}$, at which point it starts cooling efficiently. The pressure in the shocked region, $P$, is set by the shock jump conditions, so using the density profile $\rho \propto r^{-2}$, $P\propto\rho v_s^2\propto r^{-2}v_s^2\propto t^{-2}$. Using the adiabatic relation for every element $P \propto \rho^\gamma$ and the fact that behind the shock, a constant pressure implies that the product of the density and the temperature of all elements is constant, we find that $T_{line}=\Ts (P/P')^{\frac{\gamma-1}{\gamma}}$ and $t'=t\left(\frac{T_{s}}{T_{lines}}\right)^{-\frac{\gamma}{2\left(\gamma-1\right)}}$.
This gives optical/UV luminosity via line-cooling during the slow cooling regime: 
\begin{equation}\label{eq:lineas slow}
\begin{aligned}
    L_{lines,slow}&\approx L_s\cdot \frac{T_{lines}}{\Ts}\frac{t'}{t}=L_{s}\cdot\left(\frac{T_{lines}}{T_{s}}\right)^{\frac{3\gamma-2}{2\gamma-2}}\\ 
    &=4.4 \times 10^{39} {\rm ~erg/s} ~\rho_{-18}^{1.8} t_{\rm 1000}^{2.8} v_8^{2.2}\\
    &\approx 2\times 10^{37} {\rm ~erg/s} ~ \rho_{0,-19}^{1.8} v_8^{-3/2}t_{1000}^{-0.8},
\end{aligned}
\end{equation}
where we used an adiabatic index of $\gamma=5/3$.
Considering equations \eqref{eq: x slow} and \eqref{eq:lineas slow}, we find that a density $\rho_{0,-19}\simeq 10$ provides the observed luminosity in both X-ray epochs. To explain the optical emission, we need, in addition, to require $v_s\simeq 1\cdot 10^8 {\rm cm/s}$. Given the typical uncertainty in SSA analysis, this is broadly consistent with the density found by \citep{Margutti2019}.

The challenge of this model, regardless of the shock emitting the radiation, is reproducing the optical-UV broadband spectrum, which is made of line cooling emission, and perhaps some component of X-ray reprocessing in the colder regions. Naively, we \emph{do not} expect these two components to produce the observed blue spectrum, as the line cooling emission is expected to emit energy that is $\propto T$ at every temperature $T$, producing a $\nu F_\nu\propto \nu$ spectrum. Further investigation of the cooling and reprocessing spectrum is required to confirm or disprove this model. 

Regardless of the UV spectrum, this model provides a reasonable explanation for the X-rays at 220 days, by a shock with a velocity above $\gtrsim 10^8 {\rm~cm/s}$ propagating in the dilute CSM probed by the radio. If these X-rays are indeed due to CSM interaction, it implies that the dense CSM is truncated at a radius of a few $\times 10^{15} {\rm cm}$. This model, with a slightly lower CSM density at larger radii, can co-exist with any other explanation for the late UV. 

\subsubsection{Emission from a companion star}
Assuming the X-ray observation at 1350 days is unrelated to AT2018cow, and the earlier 220 day observation is due to the shock nearing the edge of the denser CSM, the observed luminosity and color in the HST observations are consistent with emission from a bright massive star. Assuming this is a companion star of the progenitor, the interaction of the ejecta with the companion would have deposited energy in its envelope. As a result, the companion luminosity would have increased immediately after the interaction with the ejecta, and then it declined gradually as its envelope radiated the excess energy. Since the observed luminosity cannot exceed the Eddington luminosity of the star, the companion mass must be $\gtrsim 10M_\odot$. 
If this is indeed the explanation, then the progenitor system should be relatively young, given the limited lifetime of the massive companion.

\subsubsection{Obscured accretion disk}
An obscured compact X-ray source has been invoked by many previous works to explain this late stage \citep[e.g.,][]{Migliori2024,Cao2024,Lazzati2024,Winter-Granic2025}. An example of such a system is an accreting compact object, which is obscured by an optically thick cloud at the required radius. The observational constraints, i.e., luminosity and colors, leave a lot of freedom for this model though they favor stellar mass compact objects  (e.g., \citealt{Winter-Granic2025}).
If the compact object is indeed a neutron star or a stellar black hole, then this accretion origin of the late-time emission is also consistent with our model for the rest of the observations. 
\subsection{Quasi-periodic oscillations}
\cite{Pasham2022} report a 3.7$\sigma$ detection of $\sim 200 \rm ~Hz$ QPOs with a root-mean-square amplitude of $\gtrsim 30\%$ in NICER X-ray observations.  \cite{Zhang2022} find a 2.2$\sigma$ detection of a $\sim 4\rm~mHz$ QPO in XMM-Newton and Swift data 6-37 days after explosion. If either of these claims is correct, the light-crossing time implied sets an upper limit to the X-ray source size of $\lesssim 10^8 {\rm~cm}$ or $\lesssim 10^{13}~\rm cm$ accordingly. Both these limits seem to be inconsistent with shock emission at $r \sim 10^{15}$ cm as the source of the X-ray, and are thus contradictory to our model. However, the significance of these detections is marginal, especially since they are estimated using a posteriori statistics, where estimating confidence accurately is practically impossible (e.g., we can never know how many QPE searches were conducted by other authors without having any detection in other astrophysical objects). If such QPOs are confirmed in the future, it will probably rule out our model.

\section{Discussion}\label{sec:discussion}
The model we propose of a fast moving ejecta that interacts with a dense CSM that covers a large fraction of the $4\pi$ sr (e.g., a geometrically and optically thick equatorial CSM) and with a lower density CSM in the rest of the $4\pi$ sr (e.g., along the poles)  explains quantitatively many of the observations of AT2018cow, using only four parameters, and is consistent with most others. 

In Table \ref{tab:disc}, we compare our model to each of the numerous observations of AT2018cow. Below, we derive properties of the ejecta and CSM that are needed to explain AT2018cow, speculate about possible progenitors, and discuss implications of applying our model to the LFBOT population.

\begin{table*}
\centering
\small
\setlength{\tabcolsep}{3pt}  

\begin{tabular}{|c|c|c|}
\hline
Observation & Relation to our model & Notes \\ \hline\hline

Bolometric luminosity &
\multirow{6}{*}{Used to constrain} &
Cannot be explained by a \\
decline rate till day 40 & & spherical CSM model, \\
& & requires a-sphericity. \\
\cline{1-1}\cline{3-3}

Time of optical peak & the model parameters & $t_{bo}$ \\
\cline{1-1}\cline{3-3}

Early photospheric radius & ($s,k,t_{bo},v_{bo}$) & Used to constrain $v_{bo}$ \\
and spectra &  & \\
\cline{1-1}\cline{3-3}

time of $L_{opt}\simeq L_{x,soft}$ & & time of $\T\simeq1$ \\
\hline\hline

Bolometric luminosity &
Explained &
$f_{\Omega}$ provides freedom \\
normalization till day 40 & & of a factor of $\sim 2$ \\
\hline

Soft X-ray spectrum &
Explained &
Fast cooling free-free is \\
& & expected to give a \\
& & $\nu^{-1/2}$ spectrum in X-rays\\
\hline

Optical emission till day 40 &
&
Originated from reprocessed X-rays,\\
& & enhanced by IC. \\
\cline{1-1}\cline{3-3}

soft X-ray/Optical ratio &
Explained &
Depends on the  \\
&(time-dependent functions) & Thomson optical depth, and on\\
& & the cooling process (free-free or IC).  \\
\cline{1-1}\cline{3-3}

Hard/soft X-ray ratio &
&
Set by the Thomson optical depth \\
& & (for a sufficiently fast-cooling shock) \\
\hline
Light curves decline at day 40&
Requires a decrease in CSM density & X-ray at 220 days suggests a steep decline in the \\
& or significant shock deceleration &   CSM density. X-ray/optical ratio predictions\\
& & are still valid (as long as the shock is fast-cooling). \\
\hline
Optical broadband spectrum &
Explained &
Reprocessed X-ray from a region\\
and NIR emission & &
 moving away from thermal\\
& & equilibrium; further modeling\\
& & required for precise predictions. \\
\hline

Broad and narrow spectral lines &
Explained &
Line widths, their appearance time,\\
& &  and their evolution are consistent \\
& &   with the velocity of the shocked material\\
& &   and upstream temperature. Further   \\
& & research required for full spectral modeling.\\
\hline
X-ray spectral feature &
Plausible & Blue shifted and broadened 6.7 or 7 keV Iron lines \\
at $\sim 8$ keV & & which disappear once the upstream cools and can absorb them \\
\hline
Radio and Sub-mm &
Explained & SSA peak explained\\
&(with freedom in & by a shock from the same\\
&  the dilute CSM density)& explosion in a lower density CSM. \\
\hline
X-ray variability &
Period and amplitude are explained&
Explained to within a \\
&to within a factor of a few & a factor of a few; 3D \\
& & simulations of the radiative \\
& & shock instability are required \\
& & for an accurate modeling \\
\hline

Late time emission &
Consistent &
\\ \hline

Optical polarization &
Broadly consistent &
The polarization is unmodeled \\
& & but broadly consistent \\
& & with asymmetric CSM. \\
\hline

X-ray QPOs &
Contradictory &
The physical size constraint \\
& (but detection is tentative) & from QPO frequency are \\
& & inconsistent with shock-powered \\
& & X-ray emission \\
\hline

\end{tabular}

\caption{A summary of the observations of AT2018cow, and how they relate to our model. All observations are discussed in \S\ref{sec:motivation}.}
\label{tab:disc}
\end{table*}

\subsection{Outflow and CSM requirements and possible progenitor systems}\label{sec:ejecta}
According to our model, immediately after the breakout, the shock becomes fast cooling. Therefore, integrating over the emission provides a good estimate of the total energy in the fast outflow that drives the shock. The integrated observed luminosity (after day 3) is $ \sim 10^{50} {\rm erg}$, and the energy emitted during the breakout peak, which we did not observe, is at most higher by a factor of a few; $E_{ej}(v\gtrsim0.1c) \sim 1-5 \times 10^{50}$ erg. Our model can also be consistent with a shock driven by a wind with luminosity declining faster than $L\propto t^{-2},~t<40\rm ~days$. The energy in this case will be dominated by the earliest time, so such a wind would not change the system energetics. 

We start by constraining the CSM. For that, we use equations \ref{eq:rho}-\ref{eq:M} for the density, CSM mass, shock velocity, and radius. We find that at day 40, the shock radius is $r_s(40d) \approx 3 \times 10^{15}$ cm, and its velocity is $v_s(40d) \approx 3,000 {\rm~km~s^{-1}}$. The X-ray emission at 220 days suggests that the shock does not decelerate much after day 40, and together with the sharp X-ray and optical drop after 40 days, they suggest that the density of the dense CSM drops significantly at $r_s(40d)$. Thus, the total mass of the dense CSM is most likely the mass collected by the shock up to day 40, $M_{CSM} \sim 0.3 {\rm~M_\odot}$. A larger mass in the CSM is possible if the late X-ray emission originates from a different source and the sharp drop in emission is due to a change in the hydrodynamic conditions, resulting in a more rapid drop in velocity. Another configuration that can accommodate a larger CSM mass is a geometrically thin high-density component, such as a thin dense disk, embedded in the dense CSM. Such a component would be enveloped by shock and thus not observed. Finally, the appearance of clear hydrogen lines, both narrow and wide, after day 15 implies that the CSM contains hydrogen at an observable level. 

Next, we constrain the ejecta that drives the shock that generates the optical and X-ray emission. Since the hydrodynamic evolution (i.e., the shock deceleration rate) rules out a reverse shock at any time, the mass of the ejecta with velocity larger than the breakout velocity must be at most comparable to the CSM mass at the breakout radius, so $m_{ej} (v_{ej} \gtrsim 0.1c) \lesssim 0.1~M_\odot$. Given that the same ejecta also drive the radio-emitting shock, it is most likely that the velocity of the ejeta that drive the shock into the dense CSM is also $v_{ej} \approx 0.1$ c and from energy considerations $m_{ej} (v=0.1c) \sim 0.01- 0.05~M_\odot$.

Finally, we place upper limits on ejecta that remain hidden (i.e., have no observable signature). The lack of reverse shock at any time until day 40 implies that the mass of ejecta as a function of velocity is limited to  $m_{ej}(>v) \lesssim 0.3 {\rm~M_\odot~}\left(\frac{v}{3 \times 10^8 {\rm~cm~s^{-1}~}}\right)^{-1/3}$. This limit is applicable down to $3,000 {\rm~km~s^{-1}}$. 
Thus, the only velocity at which a massive ($>1{\rm~M_\odot}$) ejecta can be hidden is below  $3,000 {\rm~km~s^{-1}}$. 
These stringent limits on massive ejecta most likely rule out an explosion of a star with a massive envelope, where the bulk of the mass carries most, or at least a significant fraction, of the total ejecta energy \citep{Eisenberg2022}.  They also suggest that the fast ejecta, which we see, is formed in a relatively clean environment without interacting with a lot of unobservable mass (of the order of a few solar masses), since it is difficult to deposit a large fraction of the energy in a small fraction of the total ejecta mass without putting at least a comparable amount of energy in the bulk of the ejected mass.

In principle, any progenitor that can supply these criteria is valid. The large asymmetry is consistent with a system involving an interacting binary, in which case, the dense CSM may result from rapid mass loss through L2 \citep{Lu2023,Scherbak2025}. One possible progenitor is accretion-induced collapse (AIC) of a white dwarf to a neutron star. The ejecta needed to explain the observations are within the range of theoretical predictions of the ejecta from AIC \citep[e.g.][]{Dessart2006,Sharon2020}. It is unclear, however, whether an AIC model is consistent with a $\sim 0.3 M_\odot$ CSM that contains an observable amount of hydrogen. In this regard, a single-degenerate AIC is probably more promising than a double-degenerate one, although it may be a challenge to have the needed evolution of the mass transfer rate, so it is high enough at first to have enough mass loss through L2 to make the CSM, and then slow enough to increase the mass of the white dwarf \citep{Nomoto2007}. 
If indeed AIC is the progenitor of AT2018cow, then we expect similar events to take place in both star-forming and non-star-forming environments.

Another possible progenitor is an ultra-stripped supernova (SN), which is expected to have the required ejecta mass and energy \citep[e.g.][]{Tauris2013,Mor2023}. The main challenge of this model is having a massive CSM with hydrogen. Yet, if AT2018cow is an ultra-stripped SN, similar events are expected to be detected in, or close to, star-forming regions. In either case, if some of the mass-loss remains bound in a circum-binary disk \citep[e.g.,][]{Pejcha2016}, perhaps the hydrogen in the CSM could originate from an earlier evolutionary stage.
Tidal disruption of a Helium star by a stellar mass black hole, \citealt[e.g.,]{Klencki2025} is another potential progenitor. It provides a qualitatively similar CSM geometry and similar energetics to what our model requires, and predicts that most of the energy will be released in the fast ejecta, though current models suggest CSM and ejecta masses that are larger than inferred here. Another challenge for accretion-powered progenitors is that their fast ejecta tends to be in the polar direction, while our model suggests a more isotropic source.

\subsection{Applicability to other LFBOTs}
While the model described here provides a good explanation for the multi-wavelength observations of AT2018cow, we do not expect all interaction-powered transients, even those with a very similar setup, to produce emission with similar properties. The primary reason is the diversity of phenomenologies expected from different interaction regimes. For example, a model with AT2018cow-like properties but with double the breakout velocity and a slightly shallower velocity decline would spend its Thomson thick part of the evolution in the inefficient reprocessing regime, and thus would not show a hard X-ray hump, or perhaps would show a hard X-ray hump that does not decrease monotonically (this regime requires proper modeling to be conclusive). After that, it would transition into the slow cooling regime; its bolometric luminosity evolution would show a break, and its X-ray light-curve would not show the variability. Conversely, A 2018cow-like event with half the breakout velocity would spend a significant part of its evolution in the Thomson thick fast free-free regime, and would not reach $\T=1$ until after its velocity drops below $1,000 {\rm~km/s}$ and its emission becomes line-cooling dominated. Such a system would not reach $L_{opt}=L_{x,soft}$ at any point of its evolution.

Interaction-powered transients exhibit diverse emission regimes that depend on the ordering of the four timescales (dynamical, free-free cooling, IC cooling, and Coulomb coupling), the Thomson optical depth between the shock and reprocessing region, and the X-ray luminosity's ability to photoionize the shock upstream. Most of these regimes are not well understood and require modeling before we can explore the expected range of interaction-powered transients. At this point, we cannot point to a single true and tested signature that is always observed in interaction-powered transients and can be used to clearly identify them as such. Instead, we point to several signatures, all of which can also be absent in some cases, that point towards interaction, and should alert to further investigation. (i) Narrow lines, often treated as the hallmark of interaction, can indicate interaction in many cases, but were suppressed in AT2018cow by the high temperature of the upstream. (ii) High bolometric luminosities, due to the high efficiency of converting kinetic energy into emission. This is expected, especially while the shock is fast cooling (see Figure  \ref{fig:phase_space}) (iii) A Compton hump above $\sim10\rm ~keV$ that eventually disappears. While not all systems should show a Compton hump, the disappearance of a Compton hump should point towards $\T\to 1$. (iv) A phase of joint X-ray and optical evolution with $L_{x,soft}\simeq  L_{opt}$ hints at marginally Thomson thin reprocessing. In general, $L_{x,soft}\lesssim  L_{opt}$ is expected for any situations in which the downstream has high enough column density to reprocess at least half of the X-ray emission. (v) Receding optical blackbody radius with $T_{BB}\simeq 1-2 \times 10^{4} {\rm K}$ (note that since this value depends on the cooling function, it is composition dependent) and possibly an IR excess. While not always observed, when it is, it strongly suggests reprocessing.

Full application of our model to additional LFBOTs is beyond the scope of this paper, and most of the events observed so far lack the X-ray coverage to fully test the applicability of our model. That being said, the diversity of X-ray and optical emission possible from interaction may help explain the diversity in the LFBOT population.  In addition to AT2018cow, another LFBOT with exquisite data is the recent AT2024wpp \citep{LeBaron2025,Nayana2025b}, which may be consistent with being fully interaction powered. In our framework, we interpret the peak (at $\sim 4-5$ days) as a shock breakout. The shock breakout dominates the X-ray emission until day 15-20 (note that 20 days corresponds to 3-5 times $t_{bo}$, which for 2018cow is equivalent to $t\lesssim 3-5\rm ~d$, a time till which there are very few observations). Thus, we can start applying our model around day 20. The receding blackbody radius with a temperature of $\sim 2\cdot 10^4$ at these times, and  $L_{opt}\simeq L_x$ at day $\sim 50$, hint at CSM interaction. Many of the other X-ray and optical features are different than those seen in AT2018cow; including the steep rise in the soft X-ray light-curve between 20 and 50 days and the hard X-ray evolution. Thus, if this event is powered by interaction, as some of the observables suggest, then it occupies different regimes of the phase space than AT2018cow, which must be modeled before we can test whether AT2024wpp can be fully explained in the interaction framework, and potentially constrain the system parameters. Another LFBOT with extensive X-ray coverage is AT2022tsd \citep{Matthews2023,Ho2023b}, which stands out due to its evolution with $L_x\sim 10L_{opt}$, and minute-duration optical flares. Its X-ray and optical evolution may be consistent with CSM interaction in the Thomson thin regime, but the minute-long bright optical flares \citep{Ho2023b} likely require a different energy source, such as an accreting compact object. We stress again that our model is consistent with a compact object progenitor/remnant.
\section{Conclusions}\label{sec:summary}
We present a quantitative model for AT2018cow powered by a shock propagating through an aspherical CSM. To the best of our knowledge, this is the first model of AT2018cow with significantly fewer free parameters than observables. Four hydrodynamic parameters, which fully constrain the shock evolution in the dense CSM, account for a range of observations in X-ray and optical.
The key insight is that the X-ray source is freshly shocked material just behind the shock-front itself, and the optical emission is powered by X-rays that are reprocessed by material that has crossed the shock earlier (but within the last dynamical time) and cooled down.  This geometry naturally explains two features that have challenged previous models: (i) the coordinated X-ray/optical evolution with $L_{x,soft} \simeq L_{opt}$ after day 20, and (ii) the hard X-ray hump that disappears around day 15, reflecting the transition from Thomson-thick to Thomson-thin conditions.
The velocity decline rate, pointed to by the steep light-curve decline, requires aspherical CSM. We propose an aspherical CSM with a covering factor of order unity, embedded in a more dilute CSM, such as an equatorial CSM with dilute polar regions. The combination of dense and dilute CSM regions can simultaneously explain the X-ray/optical emission from a shock in the denser CSM, and the radio observations of a $\sim$0.1c shock in lower-density CSM. The requirement for asphericity is also broadly consistent with the observed optical polarization. 
The observed X-ray fluctuations arise naturally from a radiative shock instability once the shock becomes  fast free-free cooling, though 3D simulations are needed for accurate predictions of this process.
Several additional puzzling observations find simple explanations in our framework. The NIR excess results from reprocessed X-rays in matter far from thermodynamic equilibrium, the same mechanism explains why the inferred blackbody radius decreases despite the actual shock radius increasing. The broad and narrow spectral lines originate from the shocked material and cool upstream accordingly, appearing when $\T$ drops below unity and the upstream temperature decreases enough for line formation.
Our model implies an explosion with $E_{ej}\sim 1-5\cdot10^{50}$ erg, and ejecta with mass $0.01-0.05 M_{\odot}$ at velocity $\sim$0.1c, and a CSM at a radius of $\sim 10^{15}$ cm, with a total mass of $M_{\rm CSM}\sim 0.3{\rm~M_\odot}$ and a density profile $\rho\propto r^{-2.5}$. 

The transition at $\sim$40 days, where the luminosity decline steepens but $L_{x,soft}/L_{opt}$ remains unchanged, suggests the shock reaches the edge of the dense CSM while the emission and reprocessing continue to operate as predicted. Our model is agnostic of the source of late-time UV-optical and X-ray emission. It may arise from continued interaction with dilute CSM, or from other sources, such as a companion star, or accretion onto a neutron star or a stellar mass black hole. 

Any system that can generate an explosion and CSM with these properties can potentially be the progenitor of AT2018cow. We speculate about AIC and ultrastripped SN as two possible scenarios. These events are likely to generate ejecta with the required properties, but it is not clear if any of them can have the CSM needed to explain the observations, and we will not be surprised if the progenitor turns out to be neither of these events.

This model focused on 2018cow, and the regimes needed to explain its observations. The large number of different regimes expected in interaction-powered transients, and the qualitative variation between them, predicts that 2018cow-like events will vary significantly in their phenomenology, even if their intrinsic parameters (energetics, ejecta mass, and CSM mass) vary by less than an order of magnitude. This diversity may help explain the diverse X-ray and optical emission in the LFBOT population.

Future work is needed to fully explore parameter space, and understand various components of the model, including (i) the global radiative shock instability in 3D, (ii) the reprocessing of X-ray emission in the downstream, and (iii) the hydrodynamics of radiative shocks in aspherical CSM.

\section{Acknowledgments}
 We thank Brian Metzger for his extensive comments. CMI thanks Dan Perley and Conor Omand for helpful discussions on LFBOT observations. This research was partially
supported by a consolidator ERC grant 818899 (JetNS) and an ISF grant (1995/21). This work benefited from interactions supported by the Gordon and Betty Moore Foundation through grant GBMF5076 and through interactions at the Kavli Institute for Theoretical Physics, supported by NSF PHY-2309135. KH and CMI are supported by the JST FOREST Program (JPMJFR2136) and the JSPS Grant-in-Aid for Scientific Research (20H05639, 20H00158, 23H01169, 23H04900).
\section*{Data Availability}
The data underlying this article will be shared upon reasonable request to the corresponding author.
\bibliography{refs}{}
\bibliographystyle{aasjournal}
\appendix

\section{The thin shell approximation}\label{App:thin_shell}
In principle, the profiles of the hydrodynamical quantities are set by the shock conditions. However, since the cooling time is much shorter than the dynamical time, we can find the approximate conditions by assuming a steady state, and neglecting the interaction between the mass elements between the time they crossed the shock and the time at which they are completely cooled. Under these assumptions, every point in the region over which the shock cools should obey the mass and momentum continuity equations with respect to the upstream, and these give us the profiles $\rho\left(T\right)$ in the downstream. 
Consider the shock jump conditions, where the subscript $us$ denotes the upstream properties, and the subscript $2$ denotes some point of interest in the downstream, with $v_2$ measured in the lab frame.  The shock jump conditions are:

\begin{equation}\label{eq:RK_den}
    \rho_{us}v_{s}=\rho_2 (v_s-v_{2})
\end{equation}
\begin{equation}\label{eq:RK_momentum}
p_{us}+\rho_{us}v_{s}^{2}=p_{2}+\rho_2 (v_s-v_{2})^{2}
\end{equation}

Where: $p_{us}=\frac{\rho_{us}}{m_{p}}k_{B}T_{us}$, and $p_{2}=\frac{\rho_2}{m_{p}}k_{B}T_{2}$
solving equation \eqref{eq:RK_den} for $v_s-v_2$ and plugging into \eqref{eq:RK_momentum}: 
\[
\frac{\rho_{us}}{m_{p}}k_{B}T_{us}+\rho_{us}v_{s}^{2}=\frac{\rho_2}{m_{p}}k_{B}T_{2}+v_{s}^{2}\frac{\rho^{2}{}_{us}}{\rho_2}
\]
assuming $\frac{\rho_{us}}{m_{p}}k_{B}T_{us}$ is small relative to
the other terms (which is always the case for high-mach number shocks):
\[
\frac{\rho_{us}}{\rho}v_{s}^{2}=\frac{1}{m_{p}}k_{B}T_{2}+v_{s}^{2}\frac{\rho^{2}{}_{us}}{\rho^{2}}
\]
And solving for $T_2$, we find the relation to the density:
\[
\frac{1}{m_{p}}k_{B}T_{2}=v_{s}^{2}\left(\frac{\rho_{us}}{\rho}-\frac{\rho^{2}{}_{us}}{\rho^{2}}\right)\underbrace{\simeq}_{\rho\gg\rho_{us}}v_{s}^{2}\frac{\rho_{us}}{\rho_2}
\]
This gives approximately constant pressure, which is slightly higher than in the immediate downstream in the adiabatic case. For comparison, for an adiabatic shock with $\gamma=\frac{5}{3}$,
$\frac{\rho_{ds}}{\rho_{us}}=4$, and $p_{2}=\frac{3}{4}\rho_{us}v_{s}^{2}$.

\section{Free-free spectrum}\label{App: ff_spec}
The cooling free–free spectrum, assuming a steady-state evolution (i.e., shock conditions don't change considerably over a cooling time), is identical to the spectrum obtained by placing plasma in a box, letting it cool, and summing the emission over its entire thermal history. In both cases, at each $dT$, the plasma radiates energy $\propto dT$, with a spectrum $\frac{J_{\nu,ff}(T)}{J_{ff}(T)}$, where $J_{\rm ff}\propto n^2 T^{1/2}$ and
$J_{\nu,\rm ff}\propto n^2 T^{-1/2}\exp(-h\nu/k_B T)$ are the free–free emissivity and specific emissivity (Gaunt factor assumed constant). 

The spectrum follows from integrating:
\begin{equation}\label{eq:intergral_speec}
\epsilon_\nu \propto
\int_{T_{\max}}^{0}
\frac{J_{\nu,\rm ff}(T,n)}{J_{\rm ff}(T,n)} dT
= h \int_{T_{\max}}^{0}
\frac{\exp\left(-\frac{h\nu}{k_B T}\right)}{k_B T} dT
= h\, \Gamma\left(0,\frac{h\nu}{k_B T_{\max}}\right),
\end{equation}
where the function $\Gamma(s,x)\equiv \intop_x^{\infty}u^{s-1}\exp(-u)\rm du$ is the upper incomplete gamma function of order $s$.

For $h\nu \ll k_B T_{\max}$, the integral reduces to
\begin{equation}
\epsilon_\nu \propto \log\left(\frac{k_B T_{\max}}{h\nu}\right),
\end{equation}
However, this approximation is only appropriate for $h\nu \ll kT$.

At higher frequencies, we find that an excellent approximation is:
\begin{equation}
\epsilon_\nu \propto \nu^{-1/2}\exp\left(-\frac{h\nu}{k_B T_{\max}}\right).
\end{equation}

\begin{figure}
\centering
\includegraphics[width=0.5\linewidth]{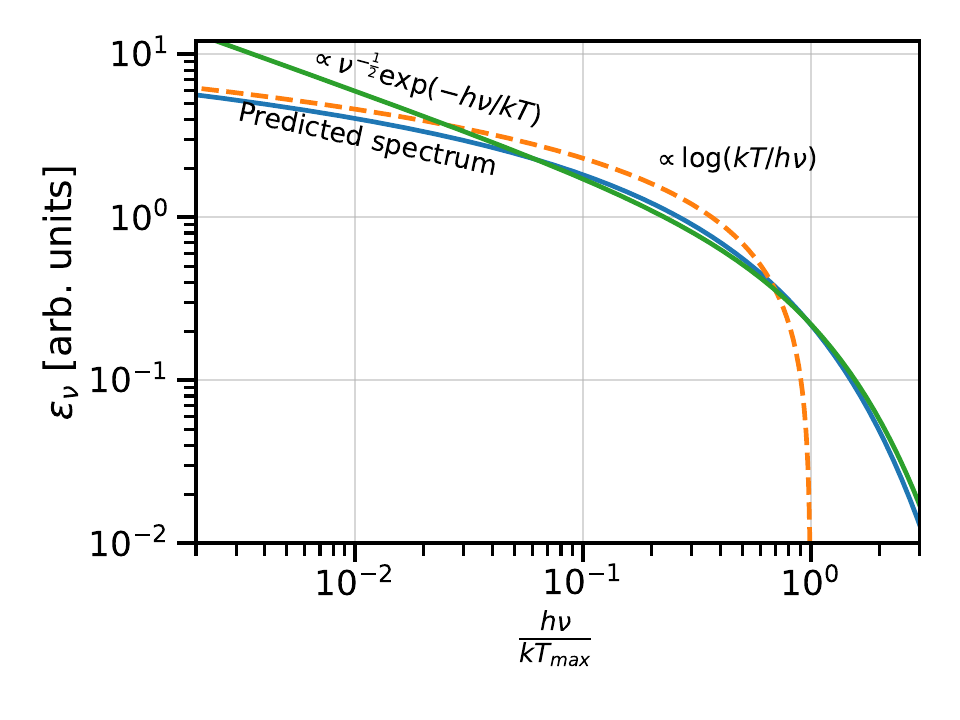}
\caption{Free–free spectrum from a cooling plasma compared to its low- and high-frequency approximations. As described in the text, $\nu^{-1/2}\exp(-h\nu/kT_{max})$ fits the predicted spectrum above for $h\nu \gtrsim 0.02 k_BT_{max}$. $\log(kT_{\max}/h\nu)$ is a good approximation at lower energies, and as expected, does not capture the high-energy cutoff.}
\label{fig:ff_spec}
\end{figure}

Figure \ref{fig:ff_spec} compares the predicted spectrum obtained from the integral in Eq. \eqref{eq:intergral_speec} with these two approximate forms. It shows that $\nu ^{-1/2}\exp(-h\nu/k_BT_{max})$ is an excellent approximation for $h\nu \gtrsim 0.02~k_BT_{max}$. We expect this to be the relevant regime for most observed cooling free-free spectra, since $T_{max}$ is either $\Ts$ (equation \ref{eq:Ts}), in the free-free dominated regime, or $T_{IC=ff}$ (equation \ref{eq:Ticff}) in the IC dominated regime. Evaluating each in the relevant regime, we find $T_{max}\lesssim 60 \rm ~keV$. On the low energy side, at around $k_BT\simeq 1\rm ~keV$, line cooling takes over, and we expect the spectrum to flatten out, as contributions from lower energies are negligible. The combination leaves at most a factor of $\sim 60$ between $T_{max}$ and the lowest temperature, making $\epsilon_\nu\propto\nu^{-1/2}\exp(-h\nu/k_BT_{max})$ an excellent approximation. 

\section{Reprocessing region in IC cooling}\label{App: reprocessing}
For there to be efficient reprocessing of the soft X-ray, the matter must become cool enough to reprocess soft X-rays within a single scattering mean-free-path from the shock front. When the upstream Thomson optical depth is below unity, so is the downstream Thomson optical depth, and this condition is satisfied even by adiabatic cooling. Here we focus on the Thomson thick regime. 
A sufficient condition is that free-free cooling takes over within $\sim\tdyn/\T$, since free-free cooling at constant pressure (which is the case here, see Appendix \ref{App: ff_spec}), cools the matter completely within a single cooling time. 
As the inverse Compton cooling rate is independent of the matter temperature, the temperature reached by IC cooling within $\sim\tdyn/\T$ is $T_{s}\cdot \exp(-\frac{\tdyn}{\tic\T})$, and we must check whether it is below $T_{IC=ff}$. We consider two cases, depending on whether the electrons and ions are in equilibrium in the immediate downstream. If they are, we can use equations \eqref{eq:tdyn},\eqref{eq:tic}, and \eqref{eq:Ticff} to find that the criterion is: 
\begin{equation}
    1.5 \T^{2/3} \vsn^{8/3} e^{-1.15 (s-1) \T  \vsn^2}\le 1
\end{equation}
Which we find holds for all $\T\ge 1$. 
If $T_e\ll T_i$, we need to, in addition, recall that the electron temperature is given by equation \eqref{eq:Ticei}, and the ion temperature is $\approx\Ts$, and find that the criterion is:
\begin{equation}
        1.5 \T^{2/3} \vsn^{8/3} e^{ -3 \Lambda_{e,25}^{2/5}(s-1) \T ^{3/5} \bar{Z}^{4/5}\vsn^{-2/5}}\le 1.
\end{equation}
For $\T>1$ is can be approximated as:
\begin{equation}
    \T\gtrsim 0.3 \vsn^{2}. 
\end{equation}
\end{document}